\documentstyle[12pt,aaspp4,psfig]{article}
\def\lsim{~\rlap{$<$}{\lower 1.0ex\hbox{$\sim$}}}
\def\gsim{~\rlap{$>$}{\lower 1.0ex\hbox{$\sim$}}}
\def\p3m{P$^3$M}

\begin{document}
\title{FUNDAMENTAL DISCRETENESS LIMITATIONS OF COSMOLOGICAL N-BODY CLUSTERING
SIMULATIONS}
\author{Randall J. Splinter\altaffilmark{1,2},
Adrian L. Melott\altaffilmark{3}, Sergei F. Shandarin\altaffilmark{3},
and Yasushi Suto\altaffilmark{4}}

\affil{randal@convex.hp.com, melott@kusmos.phsx.ukans.edu,
  sergei@kusmos.phsx.ukans.edu,  suto@phys.s.u-tokyo.ac.jp}

\begin{abstract}

  Fundamental physical considerations and past tests suggest that
  there may be a problem with discreteness error in N--body methods
  widely used in cosmological clustering studies.  This could cause
  problems with accuracy when coupled to hydrodynamics codes.
 We therefore investigate some of the effects that discreteness and
  two--body scattering may have on N-body simulations with
  ``realistic'' cosmological initial conditions.

  We use an identical subset of particles from the initial conditions
  for a $128^3$ Particle--Mesh (PM) calculation as the initial
  conditions for a variety of Particle--Particle--Particle Mesh
  (P$^3$M) and Tree code runs. The force softening length and particle
  number in the \p3m and Tree code runs are varied and results are
  compared with those of the PM run.  In particular, we investigate
the effect of mass resolution (or equivalently the mean interparticle
separation) since most ``high resolution'' codes
  only have high resolution in gravitational force, not in mass.  We
  show the evolution of a wide variety of statistical measures.
The phase-insensitive two--point statistics,
  $P(k)$ and $\xi(R)$ are  affected by the number
  of particles when the force resolution is held constant, and differ
  in different N--body codes with similar parameters and the same
  initial conditions.  Phase--sensitive statistics show greater differences.
Results converge at the mean interparticle separation
  scale of the lowest mass--resolution code.  As more particles are
  added, but the absolute scale of the force resolution is held
  constant, the \p3m and the Tree runs agree more and more strongly
  with each other and with the PM run which had the same initial
  conditions, suggesting that the time integration is converging.
  However, they do not particularly converge to a PM run which
  continued the power law fluctuations to small scales. This suggests high
particle density is necessary for correct
  time evolution, since many different results cannot all be correct.
  Our results showing the effect of the presence or absence of
  small--scale initial power suggest that leaving it out is a
  considerable source of error on comoving scales of the missing wavelengths,
  which can be resolved by putting in a high particle density.

  Since the codes never agree well on scales below the mean comoving
  interparticle separation, we find little justification to use
  results on these scales to make quantitative predictions in
  cosmology.  The range of values found for some quantities spans
  50\%, but others, such as the amount of mass in high density
  regions, can be off by a factor of three or more. Our results have strong
implications for applications such as the density of galaxy halos, early
generation objects such as QSO absorber clouds, etc.

\altaffiltext{1}{Center  for Computational  Sciences, 325  McVey Hall,
University of Kentucky, Lexington, KY 40506}

\altaffiltext{2}{Current Address: Hewlett--Packard Company, High
Performance Computing Division, 20 Perimeter Summit Blvd, MS 1904, Atlanta,
GA 30319--1417}

\altaffiltext{3}{Department  of Physics  and Astronomy, University  of
Kansas, Lawrence, KS 66045}

\altaffiltext{4}{Department of   Physics  and RESCEU,   University  of
Tokyo, Tokyo 113, Japan}

\end{abstract}


\section{Introduction}

A fundamental approximation underlying cosmological N-body simulations
is that the density field of the universe may be represented by a set
of $N$ discrete particles.  The Universe is thought to be dominated by
some form of dark matter whose mass is small compared to that of
galaxies, probably small compared to that of stars.  The formation of
structure in most scenarios proceeds from a nearly homogeneous mass
distribution with small perturbations.  The formation of stars,
galaxies, clusters, and superclusters is probably a hierarchical
process, built on a spectrum of density perturbations in this
background of very low mass particles.  In this case, the only
significant discreteness is that which emerges as the particles begin
to cluster.  The particles themselves (atoms, axions, neutrinos, WIMPS, or
something else) have such a small mass as to be insignificant compared
to that of the aggregates being studied in the simulations.  Therefore
any successful N--body code should approximate the solution to the
Poisson--Vlasov equations.  This can be coupled to hydrodynamics, to
model the baryonic matter that gave rise to the luminous component.
In this case, the gravitational force on each simulation particle
should be dominated by the mean field, with insignificant discreteness
effects (of which there are many kinds).  To trace exactly the
evolution of the {\it discrete} particle distribution may sometimes be
in conflict with this approximation. The correct evolution of the
density field, sampled by a set of particles, is supposed to be
obtained by properly adjusting the parameter $a$ which effectively
softens the gravitational force like $\propto (r^2+a^2)^{-1}$.  (There
are a variety of details of the shape of the softening of the
potential at short ranges; this is only a representative example).  If
the dark matter is not modeled as collisionless, the dominant
gravitational force on the hydrodynamic component will also be
incorrect.

The choice of $a$ is a subtle problem unresolved in practice.  If the
real universe consists of galaxies, with no internal degrees of
freedom, then $a$ could be simply set to the typical size of galaxies
$\sim 10$ kpc provided that the number density of simulation particles
$\overline n_{\rm sim}$ is comparable to that of galaxies $\overline
n_{\rm gal} \sim 0.01 h^3 {\rm Mpc}^{-3}$.  For a cosmological volume
$\sim (1000 h^{-1} {\rm Mpc})^3$, this requires that the total number
of particles is of order $10^7$, which was impossible a decade ago but
is quite feasible with current existing computational resources.  This
could appropriately simulate a Universe in which galaxy formation was
decoupled from large--scale structure, as often assumed in the 1970's
and earlier.  Our universe does not seem so simple; the prevailing
consensus is that our universe is made up of dark matter particles,
possibly with masses ranging from $m_{\rm DM} = 10^{-5}$ to $10^{12}$ eV.  If
this is the case, their number density is $3\times10^{77} ({\rm
  eV}/m_{\rm DM}) \Omega_0h^2$ Mpc$^{-3}$, where $\Omega_0$ is the
cosmological density parameter.  Then simulation particles do not
correspond to physical objects but rather provide a method to sample
the density field of dark matter in a very sparse manner.  In this
case $a$ must be chosen so as to suppress the artificial discreteness
and two-body collision effects of the simulation.  These effects are
negligible for the real universe even if the masses of large
stars are typical of the particle size.  Therefore they {\it should
  be} absolutely suppressed in simulations.  Of course the interaction between
galaxies is important, but galaxy formation is now thought to proceed
by a process of hierarchical clustering intimately interwoven with
large--scale structure formation.  When two--body effects between
galaxies are included, internal degrees of freedom must be included as
well.

This line of thought naturally indicates that $a$ should be close to
the mean simulation particle separation $\overline l_{\rm sep} \equiv
{\overline n_{\rm sim}}^{-1/3}$ since $a \ll \overline l_{\rm sep}$
would manifest undesirable discrete sampling due to the sparseness
of the density field (Melott 1981, 1990).
Hereafter we define $\epsilon = a {\overline
  n_{\rm sim}^{1/3}}$ so that $\epsilon = 1$ corresponds to this
choice.  While this is consistent with the PM simulation methodology
with one or more particles per cell (e.g. Hockney \& Eastwood 1988)
the choice of $\epsilon \lsim 0.1$ which {\it intends} to achieve
higher spatial resolution is nearly universal in P$^3$M and Tree code
simulations (e.g., Efstathiou et al. 1985; Suginohara et al.
1991). So the question is whether or not the choice $\epsilon =$ 1 is
too conservative and whether or not the choice of $\epsilon \lsim 0.1$
correctly traces the evolution of the underlying density field of dark
matter while retaining higher spatial resolution without suffering
from discreteness effects.  This question may have a different answer
depending on exactly what analysis is done.  We attempt to
answer it for the case of some statistical measures.

The answer to the above question should depend on the specific problem
the simulation is addressing.  Efstathiou \& Eastwood (1981) studied
collisionality (one effect of discreteness) by means of mass
segregation in a \p3m code with $\epsilon \lsim 0.1$ and found
significant two-body scattering.  Peebles et al. (1989) studied this
in a PM code and verified that $\epsilon \sim 1$ was necessary to
suppress collisionality.  Both these tests used reasonable
perturbation spectra but only had one diagnostic of discreteness
effects.  It is customary in computational physics to use solutions
with known characteristics to test codes before applying them to new
problems.  The shock tube is now often being used in this way for
hydro-dynamical simulations in cosmology, although no one thinks
galaxies formed this way.  Melott (1990) discussed the problem and presented
pictorial evidence that lack of mass resolution can damage results as much as
lack of force resolution. Melott et al. (1997) presented a clear case
against a use of $\epsilon < 1$ in one dimensional collapse of a plane
wave. This study was restricted to a formal test case and the precise
effect on realistic cosmological simulations is not clear.

Often it has been argued that most fluid elements in hierarchical
clustering contract, justifying the use of smaller $\epsilon$. In other words,
in this point of view,
$\epsilon$ only need be large enough to prevent two body relaxation in
collapsed regions. Kuhlman
et al. (1996) showed that even for regions of relative overdensity
$\delta > 10$ about half the small fluid elements which initially contract in
one or two directions expand in the third. Thus, due to
anisotropy at first collapse there is plenty of opportunity for scattering due
to unphysical discreteness of
the kind found in Melott et al. (1997). This would decouple results on small
scales from the initial conditions.

Another problem has to do with the use of correct initial conditions.
Absence of the initial perturbations on scales smaller than the
particle Nyquist wavelength but larger than the force resolution scale
results in inaccurate modeling of the merger history of clumps. This
cannot be compensated by improving the force resolution.  The initial
power spectrum above the particle Nyquist frequency could in principle
have a wide variety of forms.  Therefore, expecting the right answer
without putting in the correct initial conditions implies that the
final stage completely forgets that important piece of information;
this can be only approximately true.  As a test of this we generated
two PM models identical except for the presence or absence of initial
perturbations in the range $16k_f < k < 64k_f$, where $k_f$ represents
the fundamental wavenumber corresponding to the simulation box size.
Comparing the other models to the first one shows the effect of
differences in integration of initial conditions, while comparison to
the latter includes the effect of the presence of this part of the
initial conditions.

The purpose of the present paper is to quantitatively examine the
extent to which discreteness effects in simulations with $\epsilon <
1$ may change various statistical measures in models with a power
spectrum not unlike that present in the range of interest of many
viable cosmological clustering theories.  It may be somewhat
surprising that such a fundamental issue has not yet been examined in
detail in the past.  The proper comparison between PM, P$^3$M and Tree
codes (we will generically refer to \p3m and Tree codes as High Force
Low Mass Resolution codes, HFLMR, in this paper) over the dynamical
range which we will present below has been feasible for almost a
decade on supercomputers, and within reach of high--end workstations
for about five years.

An unpublished comparison test coordinated by David Weinberg in the later
1980's attacked some of these issues. It also found general agreement between
codes, with differences in detail on individual objects and on small scales.
There is no contradiction between our results and that study. A crucial
difference is our exploration of the effect of mass resolution while fixing the
force resolution.

\section{Models}

We use PM (Hockney \& Eastwood 1988; Melott 1981, 1986), AP$^3$M
(Couchman 1991), and Tree (Suginohara et al. 1991; Suto 1993) codes.
The \p3m code had adaptive smoothing turned off since we wish to
compare a standard \p3m method.  The Tree runs use the fixed smoothing
length in comoving coordinates, and we set a tolerance parameter
$\theta=0.2$ which is considerably smaller (and thus more accurate)
than conventional choices ($\theta=0.5\sim0.75$). $\theta$ merely controls how
far the tree expansion is carried, and thus the accuracy of long-range forces.

The initial power spectrum in all cases was $P(k) \propto k^{-1}$ up
to some cutoff, in most cases at the Nyquist frequency $k = 16k_f$
dictated by the runs with the fewest particles.  Realization of the
corresponding density field was generated using the Zel'dovich
approximation (Zel'dovich 1970) to perturb the particles from their
initial lattice (Doroshkevich et al. 1980).  All the models are
evolved in the Einstein -- de Sitter universe ($\Omega_0=1$).  The
comparisons were performed at three different epochs when the
nonlinear wavenumber $k_{\rm nl}$ becomes $16k_{\rm f}$, $8k_{\rm f}$,
and $4k_{\rm f}$, where $k_{\rm nl}=k_{\rm nl}(A)$ is defined as
%%%%%%%%%%%%%%%%%%%%%%%%%%%%%%%%%%%%%%%%%%%%%%%%%%%%%%%
\begin{equation}
\sigma^2(k_{\rm nl},A) = A^2 \int^{k_{\rm nl}}_0 P(k) d^3k = 1 .
\end{equation}
%%%%%%%%%%%%%%%%%%%%%%%%%%%%%%%%%%%%%%%%%%%%%%%%%%%%%%%
In the above $A$ denotes the expansion factor (unity at the initial
condition), and these values of $k_{\rm nl}$ correspond to the epochs
$A=22.36$, $42.13$, and $92.20$, respectively.  The latest moment we
studied corresponds to nonlinearity on the largest scale which does
not suffer from finite-volume boundary condition problems (Ryden \&
Gramman 1991; Kauffmann \& Melott 1992).  The specific runs we tested
and the model parameters are shown in Table 1.  We note that PM codes,
which have been extensively used in most physical applications with
large numbers of particles, are much faster than the other two types.
Thus our 128$^3$ PM runs took much less CPU time than even the 32$^3$
Tree or \p3m runs.  The typical limitation on PM runs is memory or
disk space, while CPU time is the typical limitation on \p3m or Tree
runs.  HFLMR codes usually run with $\epsilon < 1$, but we will
examine their behavior over a range in N and $\epsilon$, pushing them
toward $\epsilon = 1$ by increasing the number of particles while
keeping the absolute scale of force resolution $a$ constant.  So far
as we know, this crucial experiment has not previously been done with
HFLMR codes.

Our primary strategy is therefore to highlight the largely unexplored
mass resolution issue by varying the number of particles while keeping
$a$ constant (it is not {\it exactly} constant because in fact the
shape of the short-range softening function is different in all three
codes). Within a given code $a$ will be constant, so we can spot
trends.  Between codes softening will be of comparable size.  We can define
$r_{50}$ as the radius where the force drops to 50\% of the Newtonian value.
For the $P^3M$ code, $r_{50}=0.92a$. For the Tree code $r_{50}=0.87a$. For our
PM code, $r_{50}=0.95$ grid unit, albeit with considerable scatter in the
softened zone.

In most
cases we set $a = 1$ (in units where the box size is 128) in both the
Tree and \p3m codes, and run PM with a 128$^3$ mesh.  (The $P^3M$ runs had one
mesh cell per particle, but this is not the factor determining force
resolution there.) It is typical of
HFLMR codes to have $a$ considerably less than $\bar{l}_{sep}$ , the
mean interparticle separation over the entire simulation box.  In most
uses of PM codes. the opposite is true (although in gravitational
applications it has become customary to have $\epsilon = 1$ or $0.5$).

It is important to note that one cannot represent initial power to
higher than the Nyquist wavenumber of the particles or the mesh of the
FFT used to impose the initial conditions, whichever is worse (in fact
the latter is rarely a problem in cosmology).  Therefore most of our
runs only have initial power up to $k_{\rm c} = 16k_{\rm f}$, so that
comparisons of only the effects of divergent numerical integration can
be done. The $N = 64^3$ and 32$^3$ initial conditions are taken from a
subset of the $N = 128^3$ PM runs with the power spectrum cut off at
$k_c=16 k_f$.  In order to explore the effect of having initial
power at higher wave-numbers which is only possible with more
particles, we have one PM run with $k_{\rm c} = 64 k_{\rm f}$.  We
also have two each $P^3M$ and Tree runs with $a = 0.25$ to put some points
along the pure
force resolution axis (which has been emphasized in past tests by most users
of HFLMR codes).  These ``extra" runs have 32$^3$ particles, $\epsilon
= 0.0625$, implying $a = 0.25$.

We now have for the first time a large number of runs using 3
different codes with identical initial conditions, identical force
resolution, but varying mass resolution.  In particular the number of
particles varies widely enough to study the same physical system with
values of $\epsilon$ typical of HFLMR codes as well as to study the
same system with a PM code with similar force resolution {\it and}
matching mass resolution ($\epsilon = 1$).

It is important to note time-step limitations. Elementary principles of
numerical
stability require that no particle move more than a fraction of a softening
length $a$, or
about one--half mesh unit for $PM$, in a single time-step. This condition was
amply
enforced for all three codes.

\section{Visual Impressions}

\begin{figure}
\begin{center}
  \leavevmode\psfig{figure=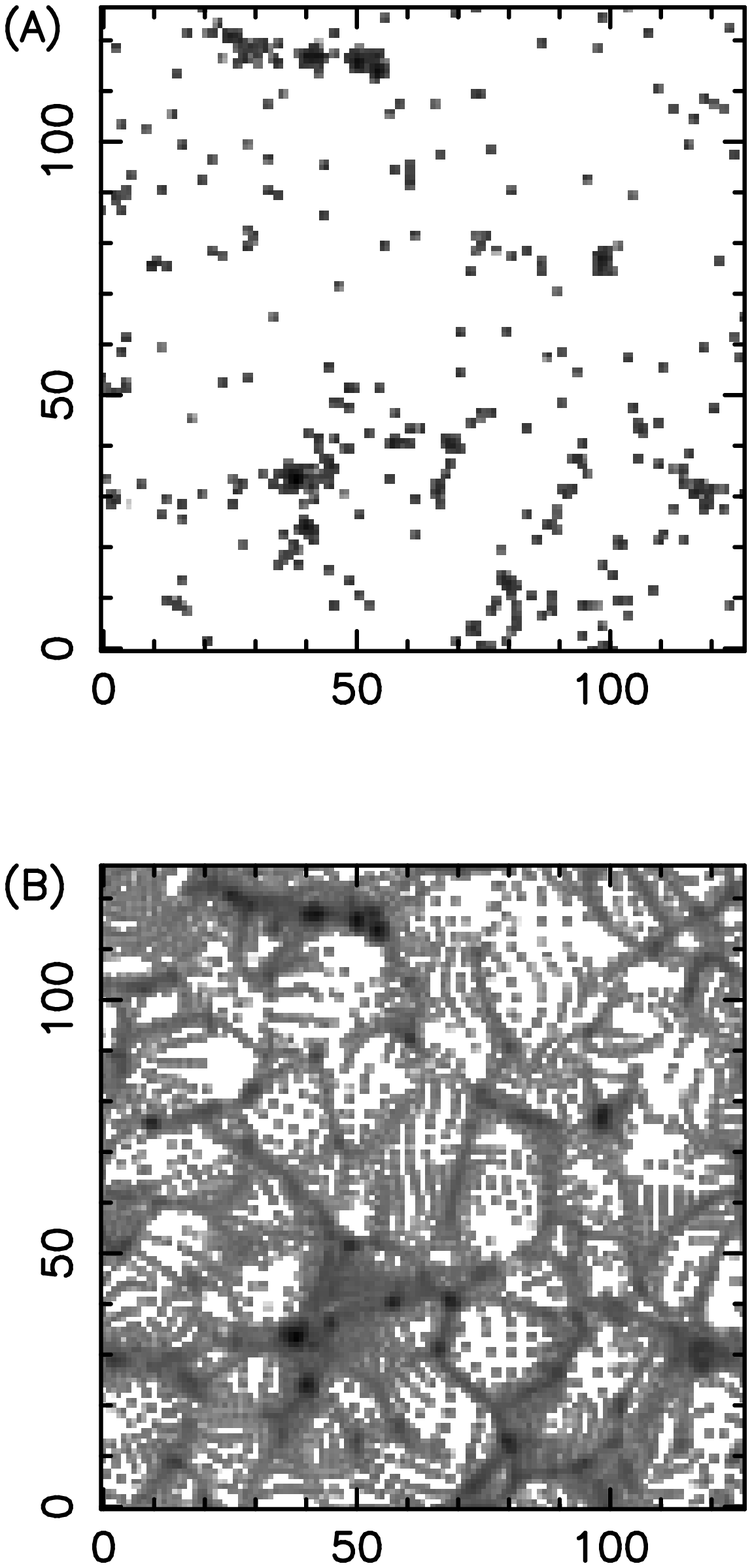,height=16cm}
\end{center}
\caption{
  We show density fields constructed from a slice of an evolved state
  of our PM simulation.  The parameters of the two images are
  identical, except that (a) is constructed using only those in the
  32$^3$ subset (see text) of particles which happen to lie within the
  slice.  Image (b) is from all of the 128$^3$ particles which lie
  within the slice.  Although the inferred two--point correlation
  would be identical, the visual impressions are very different, as
  would be the noise in higher-order measures of structure.  The general
awareness
  of superclusters which arose in the 80's is primarily the result of
  an analogous change in the signal-to-discreteness noise ratio in
  redshift surveys.
\label{fig:fig01}}
\end{figure}

The lowest order discreteness effect is merely sampling.  In
Figure~\ref{fig:fig01} we show a slice of the density field of our
PM simulation with $k_{\rm c} = 16k_{\rm f}$ at $k_{\rm nl} = 4k_{\rm
  f}$ one cell thick.  Both pictures show the same configuration; in
one case all of the particles in the slice are available and in the
other only those from a 32$^3$ subset are used.  The discreteness has
a major effect on the visual impression, corresponding to the noise
effect on high--order statistics.  A claim of filamentarity in the
sparse picture would certainly be greeted with skepticism.  Although
percolation analysis can still detect a signal in such datasets
(e.g. Melott et al. 1983) it is very noisy.  The change in
discreteness in redshift surveys is largely responsible for the shift
in attitude toward superclusters from the 1970's through the 1980's.
Clumps, filaments, and sheets are progressively harder to see through
discreteness noise; many present simulations are able to show
filaments but have too few particles to allow sheets to be seen.

More seriously, if these two different density fields were coupled to
hydrodynamics, with the dark matter driving the gravitational field,
the results would be very different. Of course, both distributions
have the same two-point correlation function, but many other things
are completely different.

\begin{figure}
\begin{center}
  \leavevmode\psfig{figure=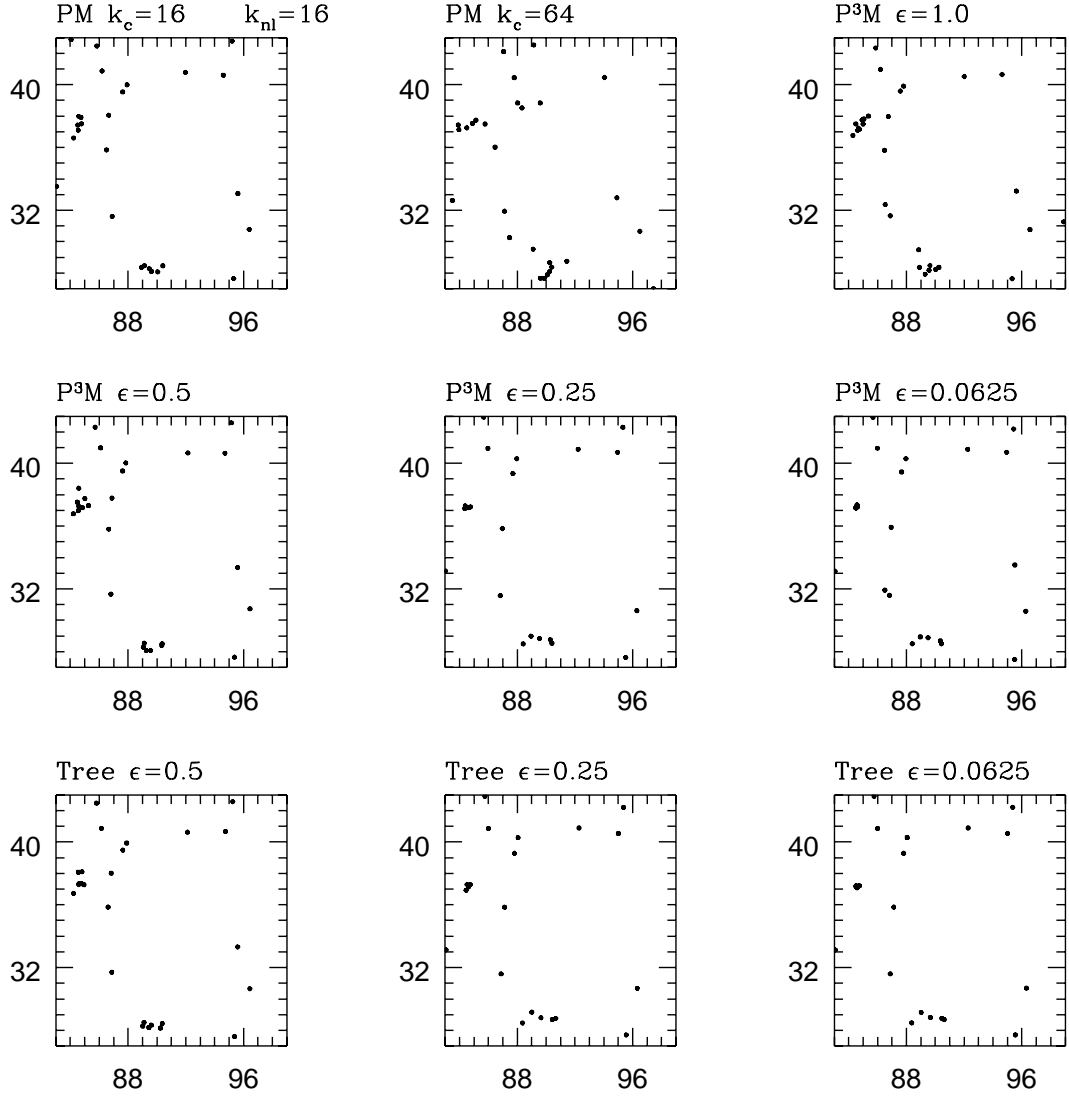,height=16cm}
\end{center}
\caption{
  For the nine runs listed in Table 1, identified by code type and by
  $\epsilon$, a slice 16 by 16 by 8 cells of a region of average
  density.  Axis units are of the 128$^3$ mesh. For the stage $k_{\rm
    nl} = 16k_{\rm f}$.
\label{fig:fig02}}
\end{figure}

Differences between the runs described in Table 1 can be seen in
Figures~\ref{fig:fig02} to 4 at different stages of evolution.  At a
given stage particles which lie within a box at the same location are
shown.  In cases with more particles, only those with the same initial
locations as those of the 32$^3$ subset are shown and used in
calculating all statistics for comparison presented hereafter.
The following statements are subjective but are made quantitative
later by cross-correlation studies.

\begin {figure}
\begin{center}
  \leavevmode\psfig{figure=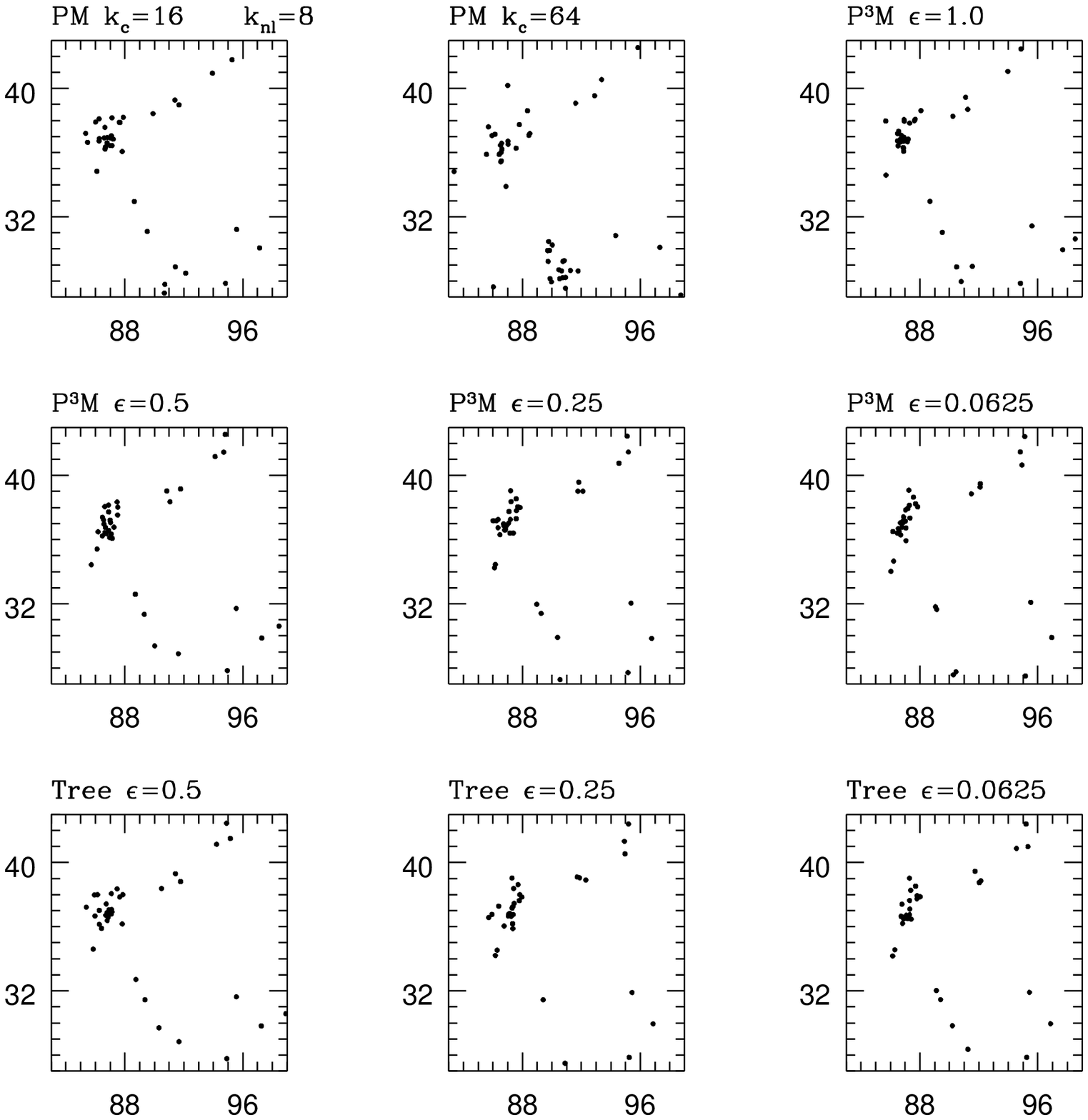,height=16cm}
\end{center}
\caption{
  As in 2, for the stage $k_{\rm nl} = 8k_{\rm f}$.
\label{fig:fig03}}
\end{figure}

\begin {figure}
\begin{center}
  \leavevmode\psfig{figure=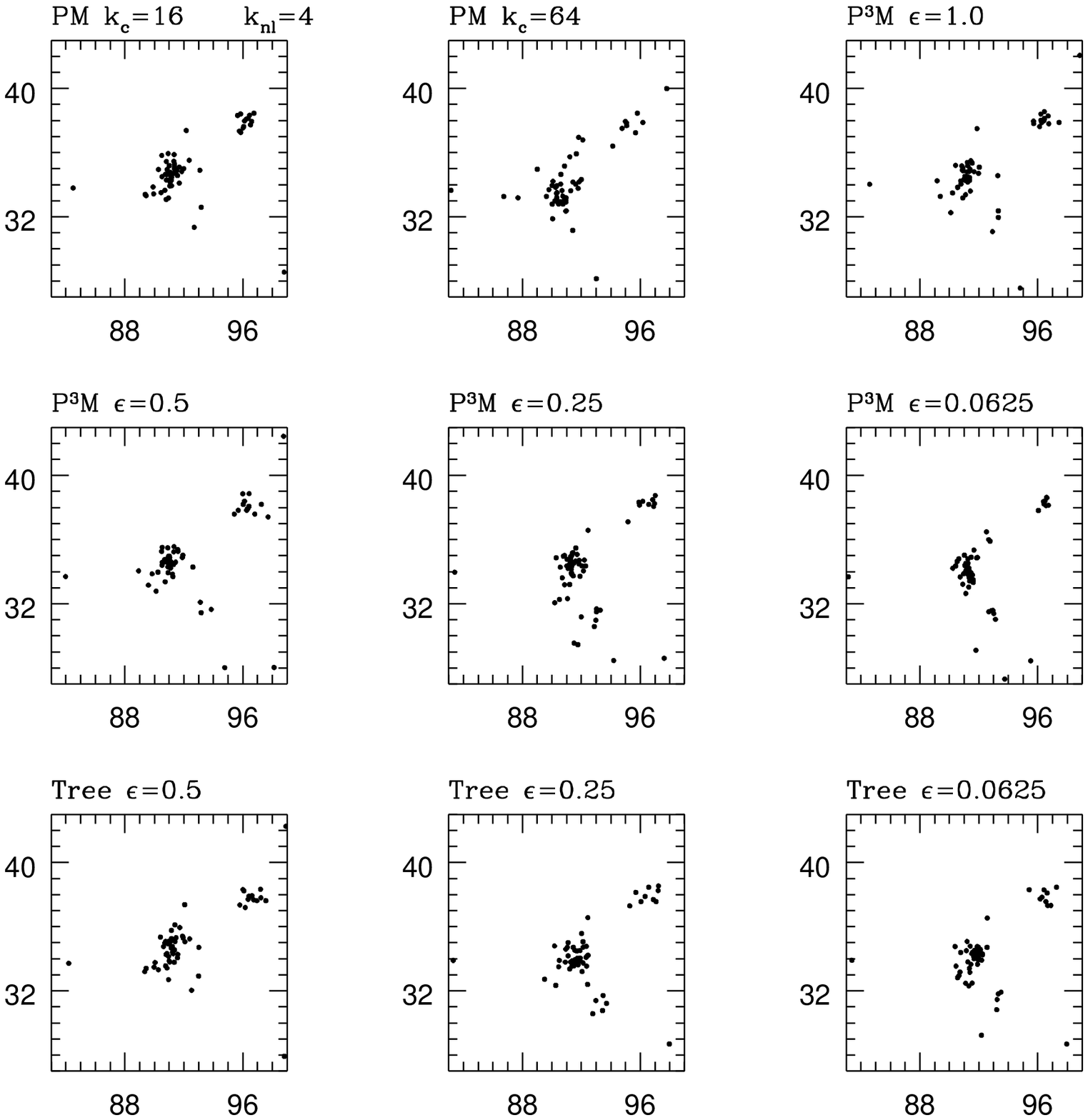,height=16cm}
\end{center}
\caption{
  As in 2, for the stage $k_{\rm nl} = 4k_{\rm f}$.
\label{fig:fig04}}
\end{figure}

There is a general resemblance of all plots at the same stage.  Since
they all have force resolution $a \lsim 1$ if this were the only issue
we would expect them to be identical down to the scale of one tick
mark.  They are not. For example, the three \p3m runs all with $a = 1$
($\epsilon$ = 1, 0.5, 0.25 corresponding to $N = 128^3, 64^3, 32^3$)
all look different on scales larger than $a$.  Within this series, as
the number of particles is increased they come to resemble the PM run
with $k_{\rm c} = 16k_{\rm f}$, the same initial condition.  The same
trend is evident in the Tree runs.  Another observation one can make
is that on a scale of 4 cells (the mean interparticle separation of
the sparsest runs) there are very few apparent differences.  The PM
run with $k_{\rm c} = 64 k_{\rm f}$ looks different from the others,
including the PM run with $k_c = 16k_f$. The overall
configuration is similar, but there are many small--scale features
reproduced by none of the other models.  Consider for example the
shape and orientation of the two
largest objects in Fig~\ref{fig:fig02}; the existence of the second
object in Fig~\ref{fig:fig03}; the partial bridge around (93,36) in
Fig~\ref{fig:fig04}.  Power on scales above the force resolution but
below the mass resolution of HFLMR has made an apparent difference. However it
is described, it is evident that the sub-panels in each of these Figures are
not
identical.
Again, on scales of the largest interparticle separation (4) the differences
appear to evaporate.
HFLMR codes are based on the idea that one can shrink $\epsilon$
far below unity to get higher resolution.  We see very real differences
implying integration errors for even the modest $\epsilon = 0.25$.  We
expect greater differences for smaller $\epsilon$.  The transfer of
power to small scales is strong, but does not make small-scale initial
power totally irrelevant.

The reader is referred to the center and
bottom center images of Figure 7 in Beacom et al. (1991) for another
example of the importance of initial power on small scales even when
it is deep in the nonlinear regime. In that paper, as well as Melott et al.
(1993), we emphasized the effectiveness of the transfer of power from long to
short waves. The general position and orientation of objects is
determined by initial perturbations on that comoving scale and larger, so
smaller perturbations are ignorable for this purpose. See also
Melott et al. (1990), Little et al.
(1991), Evrard and Crone (1992). However, here we stress that the {\it internal
structure} of these objects will vary depending on smaller--scale
perturbations. If we wish to study that internal structure, the smaller--scale
initial perturbations  must be
present and properly evolved.

In all of the papers cited in the previous paragraph, it was concluded that:
(a) There is a strong transfer of fluctuations from large to smaller comoving
scales. (b) There is a nearly negligible transfer from small to large scales.
This is the reason for the success of approaches such as the Truncated
Zel'dovich Approximation (Coles et al. 1993) which ignore initial small--
scale fluctuations, successfully follow large--scale fluctuations into the
mildly nonlinear regime, producing remarkably accurate results. Although
it was not emphasized in any of these papers, visual inspection will verify
that different small comoving--scale perturbations produce different results
on these scales. This is easiest to see in Beacom et al.(1991) due to the use
of a color scale, so that high density regions can be inspected (while they
are solid black in the typical N--body dot plots).  However, the result can
be verified by close inspection of any of them.  Small-scale power is diluted
by mode coupling, but it is not lost.

In the following sections we quantify some of the differences we found with a
variety of statistical measures.

\section{Statistical Comparisons}

This section contains the main body of our results.  Although we
examine a number of measures, this must be considered a preliminary
investigation.  We wish to stress that similarity or lack thereof in
the studies of one characteristic between various runs should not be
construed as implying the same thing for some other purpose.  The
major phase differences we find later make simple extrapolation based
on power amplitudes highly questionable.

Errors are not shown on the quantities presented herein. This is a choice
grounded in our computational/comparison strategy, which brings out {\it
systematic}
differences. All but one of our runs of different code types had {\it
identical} initial conditions. A 128$^3$
particle ensemble (or a $32^3$ or $64^3$ subset in those cases where the
Nyquist Theorem would not be violated) is evolved from this state. Comparisons
are done between a $32^3$ subset which is the same in all cases. We have found
that using a different subset of $32^3$ particles, or the full ensemble if
larger,
produces (with one exception) differences so small as to be invisible on our
plots. Thus our
sampling error is essentially zero. It would be possible to create some error
bars by bootstrap, but they would be meaningless.

Errors could be created based on cosmic variance--doing multiple realizations
of each box. This, too is essentially meaningless here. If the box size
were made larger, the cosmic variance would change, but the differences between
codes remain (since we are only adding linear modes, which almost any code can
handle). We could shrink or expand the errors at will by this linear procedure,
but
the systematic differences we find are in the nonlinear behavior.

There are two senses in which error is meaningful here:

(1) Do the differences we find matter in practice? This depends on the desired
accuracy. A
factor of two was formerly unimportant in many applications in cosmology, but
is now insufficient for many applications.

(2) What about other initial conditions? The most meaningful kind of comparison
would be to have runs with other initial power spectra, or non--Gaussian
initial
conditions. This would bring out how much the {\it systematic} differences we
find vary depending on the kind of initial conditions. This must be left for
future work.

\subsection{Power Spectra}

We compute our power spectra on a 128$^3$ mesh with CIC cloud
assignment of local density.  In all cases we base our analysis on 32$^3$ of
the
particles, so that the discreteness contribution is exactly the
same. We have not subtracted off the discreteness contribution from
the spectra, because it is not Poissonian at early times and
subtraction of such would constitute an error.  For consistency we
also do not subtract it at late times (when it would in any case be
quite small).

\begin{figure}
\begin{center}
  \leavevmode\psfig{figure=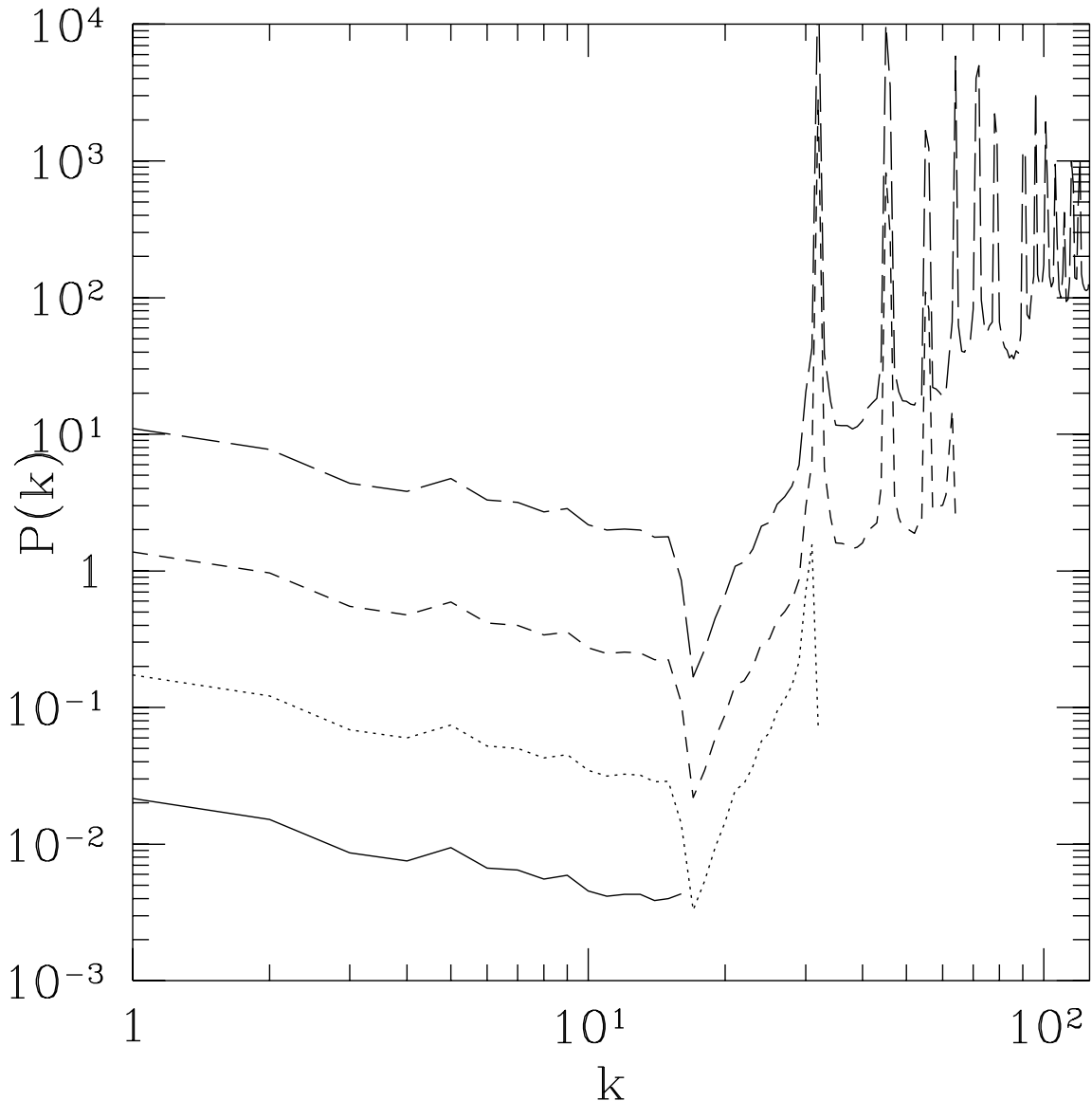,height=16cm}
\end{center}
\caption{
  The power spectrum constructed using the 32$^3$ particles of the
  initial conditions for all except our $k_{\rm c} = 64k_{\rm f}$
  runs, as evaluated on mesh of size 32$^3$ (solid), 64$^3$ (dots),
  128$^3$ (shortdash), and 256$^3$ (longdash), shown with vertical
  offset.  The normalization is such that a Poisson distribution of
  points of the same number of particles as mesh cells on each mesh
  would converge to $P(k) = 1$.  Normally, such spectra are only shown
  up to the particle Nyquist frequency, as in the solid line.  The
  spikes are a result of the lattice of particles which is deformed to
  provide the initial conditions, and are of course not random
  phase. In HFLMR codes, this part of the initial conditions is
  evolved, with unknown consequences.
\label{fig:fig05}}
\end{figure}

In Figure~\ref{fig:fig05} we show the spectrum of the initial
conditions evaluated on a series of progressively finer meshes (with a
vertical offset for clarity).  The
lowest line corresponds to the standard choice of showing things up to
the particle Nyquist frequency.  However, it is common practice at
very late times to show the autocorrelation function to very small
radii.  This seems to be a contradiction.  {\it If} the initial
spectra had been shown to the same resolution, they would look like
these (our example is from 32$^3$ particles; with more particles the spikes
lower
in amplitude and move to higher $k$).  Our point is that {\it
  if} the code has sufficient force resolution to resolve these spikes
(which are not random phase), then those {\it are} the initial
conditions.  This study then includes both an examination of
integration errors as a function of $\epsilon$ as well as the effect
of the absence of this initial power-- two independent but practically
intertwined effects.

\begin{figure}
\begin{center}
  \leavevmode\psfig{figure=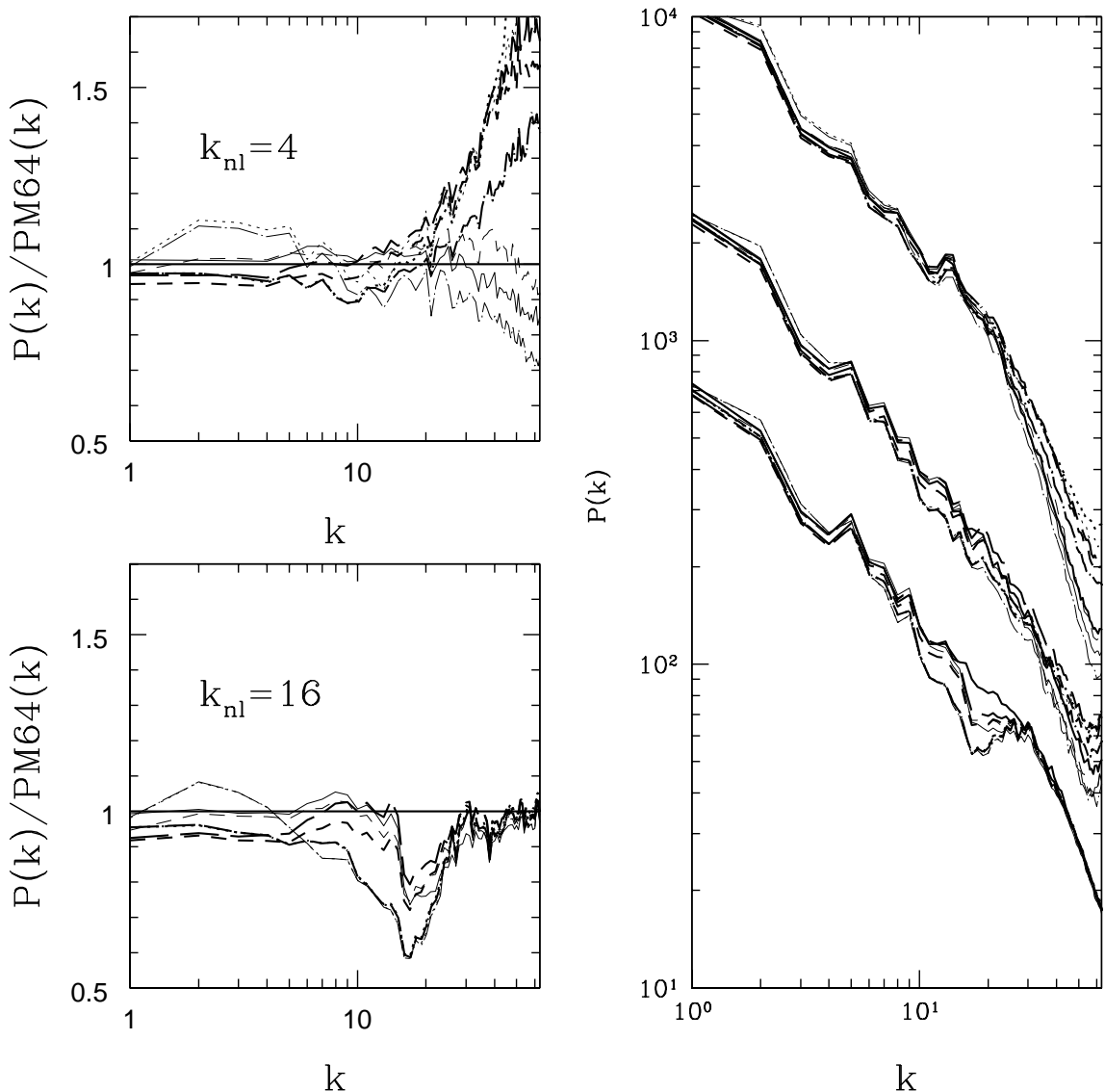,height=16cm}
\end{center}
\caption{
  Right panel:
  The power spectrum of three successive
  evolved stages
  for all our simulations, evaluated from 32$^3$ particles on a
  128$^3$ mesh (their force resolution).  The normalization is such that
  a Poisson distribution of 128$^3$ particles would converge to $P=1$
 Left panel: The ratio of the power in a given model to that in our
 fiducial $k_{\rm c} = 64k_{\rm f}$ PM run at the first (bottom)
 and last (top) evolved stage.
  The light solid line is the $k_{\rm c} = 16k_{\rm f}$ PM run;
  the heavy solid line is the $k_{\rm c} = 64k_{\rm f}$ PM run.  Other
  heavy lines are \p3m runs and light lines are tree code runs.
  Dotted lines: $\epsilon=0.0625$ runs with 32$^3$ particles.
  Longdash-dot: $\epsilon=0.25$ runs with 32$^3$ particles.
  Shortdash: $\epsilon=0.5$ runs with 64$^3$ particles.  Longdash:
  $\epsilon=1.0$ run (\p3m only) with 128$^3$ particles.
\label{fig:fig06}}
\end{figure}

In Figure~\ref{fig:fig06} are plotted all of the power spectra
from the 32$^3$ subset of the runs specified in Table 1 at our three
output stages.  We also show power ratios to help clarify differences.
Considering the diversity of types of codes and their
operating parameters, the agreement is encouraging.  In the final
output stage, major differences are only present beyond about half the
mesh Nyquist frequency, i.e., $k = 32k_{\rm f}$.  Some expected trends
emerge: the codes with the very smallest $a$ have the highest
amplitude at high frequencies for a given code type.  PM codes in
general have low amplitude at large $k$.  Other results are
unexpected: Tree codes also have systematically lower amplitude than
\p3m codes with the same $\epsilon$ at large $k$.  For both Tree and
\p3m codes, for {\it fixed} $a$, as the number of particles increases,
the nonlinear amplitude increases.  This was not anticipated.  We speculate
that
a larger number of particles is better able to handle the mode
coupling that accompanies nonlinear evolution.

Some support for this idea can be found in the neighborhood of
the ``kink" around $k = 16k_{\rm f}$ in the earliest stage shown (bottom group)
in
some evolved spectra in
Fig~\ref{fig:fig06} (we temporarily ignore the run with
additional high-$k$ power).  All the runs which had the same
initial conditions fall into two behavior groups here: those with
32$^3$ particles (dotted or longdash-dot lines), and those with more particles,
without regard to code type, mass or force resolution.  All runs with
32$^3$ particles overlie one another, and have the lowest amplitude in
the region of the kink.  Recalling that $k = 16k_{\rm f}$ is the
Nyquist frequency of the particles, it is entirely reasonable to
suppose that this kink is caused by the inability to transfer power to
higher frequencies limited by the Nyquist Theorem. Note that runs with more
particles, but no initial power beyond $16k_f$ do not show this kink.

This kink is hidden later; no doubt mode coupling from even longer
waves coupled with the considerable deformation of the lattice allows
this to fill in.  But for some codes--those with fewer particles--a
crucial step in the evolution was handled incorrectly.  The fact that
the power spectrum recovers does not mean the evolution is correct
(see also Bouchet et. al 1985).  It probably {\it does} mean one
can approximate the final power spectrum to fairly high wave-numbers with
poor mass resolution.  Note that the point of divergence of the spectra in
the evolved state is close to the particle Nyquist wavenumber ($k =
16k_{\rm f}$).  It is tempting to conclude that although spectral
agreement is good, one can only have full confidence up to this value.
Knowing whether this is a general feature or an accident will require further
study.

\subsection{Two-point Correlation Function}

For a long time the two-point correlation function $\xi(R)$ has been a
staple of cosmology studies in large-scale structure.  See Peebles
(1980) for an excellent exposition.  Formally, it is the Fourier
transform of the power spectrum discussed in the last section, so no
new information is present.  However, it is packaged differently, and
may allow new insights through new presentation.

\begin{figure}
\begin{center}
  \leavevmode\psfig{figure=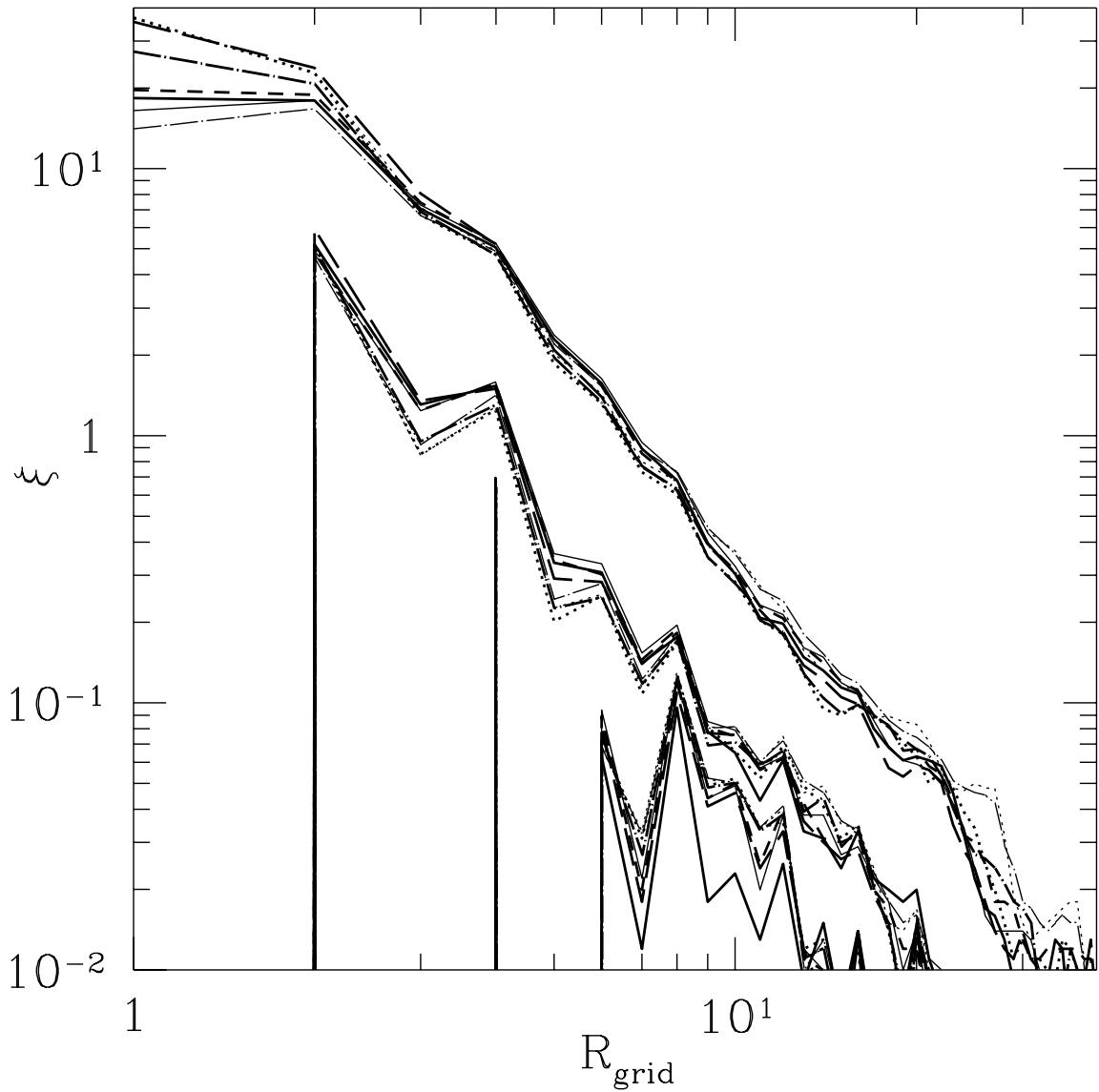,height=16cm}
\end{center}
\caption{
  The two-point correlation function of all the mass in the
  simulations is shown for the same three stages shown in the power
  spectrum plot.  Line types here and in all subsequent plots match
  Fig. 6.  The spikes which exist at early stages are a relic of the
  lattice upon which the particles began, which is insufficiently
  deformed to be suppressed then.  The last stage corresponds to
  nonlinearity sufficiently strong that boundary conditions would
  become a problem if the simulation were continued further.
\label{fig:fig07}}
\end{figure}

In Figure~\ref{fig:fig07} we show the evolution of $\xi(R)$ at our
three stages plotted against separation $R$ in grid cells (based on
128$^3$ as the box size). $\xi$ is computed (not estimated) by using Fourier
Transform techniques on a very fine ($256^3$) density mesh. Thus the
flattening shown at $R=1$ on Figure 7 is due to softening in the dynamical
force law, not any smoothing due to our method of computing $\xi$. This method
is equivalent to using all possible pairs (except those at $R<0.5)$.
We show the spikes which exist at early
stages of evolution from a lattice; they are normally suppressed in
publication comparing with data by only showing late stages.

Note that differences at the latest stage are not too great.  However,
there is a noticeable effect at the second stage near $R = 3$.  All
the codes with 32$^3$ particles ($P^3M$ or Tree) differ from the rest here. We
stress
that {\it all} correlations and spectra are computed using CIC
weighting (Hockney \& Eastwood 1988) from a set of 32$^3$ particles,
even when the original contained many more. Thus this is a systematic
difference and not an artifact of sampling.

In the final stage, while there is good agreement at larger radii, for $R <
2$ or twice the force resolution scale of most of our runs, things
begin to break down severely.  Both dotted lines ($\epsilon = 0.0625$)
are high, consistent with the traditional HFLMR emphasis on force
resolution.  However, the \p3m code with 128$^3$ particles and
$\epsilon = 1$ is also high.  We find evidence that both better mass
resolution and better force resolution contribute to maintaining the
power law to small radii, which provides support for the use of HFLMR
codes to get these statistics without good mass resolution.  However,
we do not know whether or not it is accidental.  Strangely, the \p3m
run with 64$^3$ particles and $\epsilon = 0.5$ has the lowest
amplitude at small $R$, occupying a kind of minimum.  As we do not
have a 128$^3$ tree run, we cannot be sure of the position of the
minimum there.  Still, its run with $\epsilon = 0.0625$, $a=0.25$
overlies the \p3m run.  At any rate, the non--monotonic behavior of
the \p3m run and the observed effect of both force and mass resolution
separately varied as well as code differences make it difficult to
justify the use of $\xi$ on scales below about 2$a$.  More study of
these effects is needed.  We have no way of knowing whether our
results on force or mass resolution would continue to change beyond
the range in N and $a$ studied here.  Furthermore, without further
study we cannot be sure whether the scale of agreement on $\xi$ is
coupled to the force resolution scale, the mass resolution scale, or
some complicated combination of the two.
We do not consider the differences at large radii to be very significant.
They are of low absolute amplitude, and at the latest stage they are on
a large enough scale to be suspicious on grounds of boundary conditions
(Kauffmann and Melott 1992). They may reflect some innate differences in
the Green function between the codes; the Tree code has a very different
strategy.

Conclusions on the two-point correlation: (1) $\xi(R)$ appears to
agree to within 15\% down to $R \sim 2a$, (2) Runs do not converge at smaller
separations, (3) It appears that HFLMR may approximate effects of high
mass resolution reasonably well, insofar as $\xi(R)$ is concerned, and
(4) $\xi$ in N-body simulations for $R \lsim 2a$ cannot be regarded as
totally reliable, since it changes when either $N$ or $a$ is varied
independently.

\subsection{Pairwise Velocities}

The statistics which have been discussed up to this point are based
purely upon the spatial distribution of the mass.  To probe any
differences in the dynamics we also need to consider the velocity
field. To compare the evolved particle velocities we compute the mean
pairwise (PW) velocity and the mean pairwise dispersion (Peebles
1980). We compute the mean pairwise velocity using
%%%%%%%%%%%%%%%%%%%%%%%%%%%%%%%%%%%%%%%%%%%%%%%%%%%%%%%%%%%%%%%%%%%
\begin{equation}
<V_{12}(r)> = { 1\over N(r)}
\sum_i \sum_{j \neq i}
{(\vec{v_i} - \vec{v_j}) \cdot (\vec{r_i} - \vec{r_j}) \over
 | \vec{r_j} - \vec{r_j} |}
\end{equation}
%%%%%%%%%%%%%%%%%%%%%%%%%%%%%%%%%%%%%%%%%%%%%%%%%%%%%%%%%%%%%%%%%%%
where the summation is taken over all the pairs with separation $r$
and $N(r)$ the number of such pairs.  The PW dispersion,
$\sigma_{V_{12}}(r)$, is similarly defined by
%%%%%%%%%%%%%%%%%%%%%%%%%%%%%%%%%%%%%%%%%%%%%%%%%%%%%%%%%%%%%%%%%%%
\begin{equation}
\sigma_{V_{12}}(r) = \sqrt{
{1\over N(r)}
\sum_i \sum_{j \neq i}
\left|{(\vec{v_i} - \vec{v_j}) \cdot (\vec{r_i} - \vec{r_j})} \over
{ | \vec{r_j} - \vec{r_j} |} \right|^2.
}
\end{equation}
%%%%%%%%%%%%%%%%%%%%%%%%%%%%%%%%%%%%%%%%%%%%%%%%%%%%%%%%%%%%%%%%%%%

\begin{figure}
\begin{center}
  \leavevmode\psfig{figure=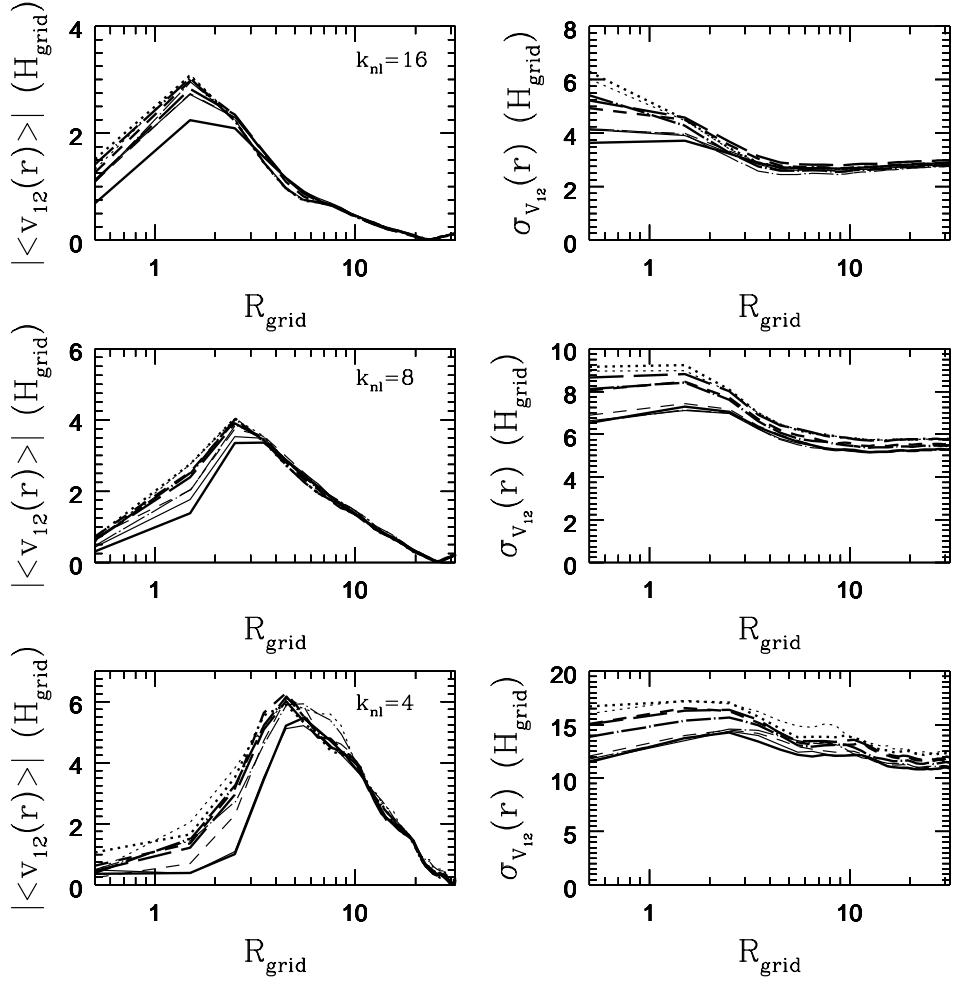,height=16cm}
\end{center}
\caption{
  The pairwise velocity and pairwise dispersion, as defined in the
  text in Section 4.4 are plotted for each stage, using the usual line
  types.  The force resolution scale is 1 for most of our runs, 0.25
  for two.  The mean interparticle separation of the sparsest runs is
  4; convergence is not reached until approximately this scale (rather
  than the force resolution scale).
\label{fig:fig08}}
\end{figure}

In Fig~\ref{fig:fig08} we plot the absolute magnitude of the mean PW
velocity and the PW dispersion.  There are several clear trends.  As
expected, when $\epsilon$ decreases both the PW velocity and
dispersion increase. This is because the particles can now approach
one another more closely and scatter more strongly off from one
another. This trend is also explained in terms of the cosmic virial
theorem incorporating the finite size effect of particles (Suto \&
Jing 1997).  There is a more interesting and unexpected trend; as the
number of particles is increased (keeping the absolute force
resolution fixed) the mean PW velocity and PW dispersion tends to
increase. While the above trend with $\epsilon$ might be expected this
trend which depends upon the mass resolution is surprising and
substantiates our claim that decreasing the force softening length
past the mean particle separation can lead to spurious effects.

\subsection{Density Cross--Correlation}

The density cross--correlation
%%%%%%%%%%%%%%%%%%%%%%%%%%%%%%%%%%%%%%%%%%%%%%%%%%%%%%%%%%%%%%%%%%%
\begin{equation}
K = {<\delta_1 \delta_2> \over {\sigma_1 \sigma_2}}
\end{equation}
%%%%%%%%%%%%%%%%%%%%%%%%%%%%%%%%%%%%%%%%%%%%%%%%%%%%%%%%%%%%%%%%%%%
was introduced by Coles et al. (1993) as a method to help quantify
similarity (or lack thereof) in the density distribution between
various models, where in our case $\delta$ is the density contrast on
the (32$^3$ or 128$^3$) mesh and $\sigma$ is its $rms$ value.  We
compute the cross-correlation on two different sized meshes. One with
$32^3$ cells which corresponds to the mass resolution scale of our
lowest mass resolution runs and another with $128^3$ cells which
corresponds to the force resolution scale of our fiducial models. This
comparison will help quantify whether the variety of models tested
here produce similar or different small--scale structure.

In Tables 2, 3, and 4 we show the density cross-correlations at the
nonlinear stages $k_{\rm nl}=16k_{\rm f}$, $8k_{\rm f}$, and $4k_{\rm
  f}$. The table entries above the diagonal are evaluated on a $128^3$
mesh and the values below the diagonal on a $32^3$ mesh.  In order to
assure uniform treatment, all density fields are computed from a set
of $32^3$ particles which had the same initial conditions.  Therefore,
any differences are due to systematic differences between the
configurations in the evolved state.  We ignore here the additional
differences introduced purely by sampling effects, as in
Figure~\ref{fig:fig01}.  However, we ran tests to insure that
densities computed from full ensembles of particles do not give
greatly different results, by comparing with full ensembles when they exist.

The cross--correlations on the $32^3$ mesh are much better than the
cross--correlations on the 128$^3$ mesh.
In fact, most results are close to unity. Thus, the codes agree rather
well at the mean interparticle separation scale of the coarsest mass
resolution runs.  This coarse-mesh agreement is generally quite high
at all stages for all codes with two exceptions: (1) The Tree code
runs with 32$^3$ particles and (2) the PM runs with a full range of
initial power $k_{\rm c} = 64 k_{\rm f}$ evolve values of $K$ closer
to 0.9 than to 1.0.  The first may be due to some large-scale
differences due to the absence of a grid in the Tree code while PM and
\p3m codes share the same grid-based algorithm in computing long-range force.
The PM results indicate that the absence of correct initial conditions
on small scales can still be detected, even at late times and on larger
scales, when those modes have gone deeply nonlinear.  These
differences, however, are not large.  Even the worst agreement on this
mesh is better than the best agreement on the fine (128$^3$) mesh
corresponding to the force resolution of most of the runs.

The fine-mesh cross-correlations (shown above the diagonal in Tables
2-4) display a number of interesting patterns: (1) Agreement between
codes with the same initial conditions, $N$, and $\epsilon$
monotonically worsens, reflecting amplification of differences.  (2)
This worsening is much more rapid between runs with the same {\it
  small} $\epsilon$.  For example $\epsilon = 0.0625$ Tree cross \p3m
(same $\epsilon$) evolves from 0.95 to 0.63 while for $\epsilon =
0.25$ it goes from 0.96 to 0.70.  For the $N = 64^3$ runs with
$\epsilon = 0.5$, it goes from 0.90 to 0.84.  This is strong evidence
that integration errors are being much more strongly amplified for
small $\epsilon$, as suggested by Suisalu \& Saar (1995), and by Park
(1997).  As emphasized by these authors, this suggests real problems for HFLMR
codes.
(3) At a given stage, as $N$ and $\epsilon$ are increased,
runs with the same initial conditions converge to one another.  PM,
Tree, and \p3m all agree rather well in this limit. There are also two
islands of consistency in Table 4 between the two \p3m runs with
smallest $\epsilon$ and $a$ (0.86), and the two Tree runs with
smallest $\epsilon$ and $a$ (0.87), but this agreement does not extent
to agreement between two {\it different codes} with small $\epsilon$
and $a$.  The convergence for large $N$ and $\epsilon$ {\it does}
extend to agreement between different codes.  (4) The run with $k_{\rm
  c} = 64k_{\rm f}$ is the best standard of comparison for appropriate
initial conditions in cosmology, since most realistic scenarios have power
extending to very small scales, not a cutoff. While the comparisons
discussed in (1-3) above highlight differences in integration, (4)
includes the effect of the presence or absence of small-scale power.
It is only possible to put this power in when a very large $N$ permits
it to be sampled by the particles.  Time evolution shows that
cross-correlation of the full-sampled run with other runs becomes
slowly better, as mode coupling brings down power from larger scales;
however, it never really achieves a high value with any other run,
regardless of the force resolution of that run.  In fact, the runs
with smallest $a$ have the {\it worst} agreement with this
full-sampled run, even though the strategy of small $a$ is to try to
follow clustering on small scales.  These results combined with those
on the correlation function/power spectrum appear to indicate that
this strategy may help produce the correct amplitudes of Fourier
components, but make the phases worse.  Thus it may be that only statistics
like the correlation function and the power spectrum which contain no
phase information show strong agreements between the various
models. Since the differences on small scales lie in phase
information, we will look at that in the next section.

\subsection{Phase Correlations}

Many different configurations can have the same two--point correlation
function (or equivalently the same power spectrum) but look completely
different.  This is due to the difference of the phases of the Fourier
component of density field which $\xi(r)$ and $P(k)$ ignore.
Gaussian-random density fields have statistical properties specified
only by amplitudes (since random phases have no information), but the
properties of non--Gaussian distributions are dependent on phases.
The end results of N--body simulations and the distribution of matter
in the Universe are both non--Gaussian, although they may have begun
from Gaussian perturbations (as our simulation did; see also
Suginohara \& Suto 1991; Lahav et al. 1993).

Testing for phase agreement is another way of checking for agreement
between density fields -- in this case those produced by the N-body
code.  Given phase information for complex coefficients $re^{i \theta}$
specified we use $\theta = \theta_a - \theta_b$ to specify the
difference in phase between two coefficients.  Our measure of phase
agreement is $<\cos \theta>$, where the averaging is over spherical
shells in $k$.  This measure is $1$ for perfect correlation, $0$ for
uncorrelated distributions, and negative for anti-correlated ones.

\begin{figure}
\begin{center}
  \leavevmode\psfig{figure=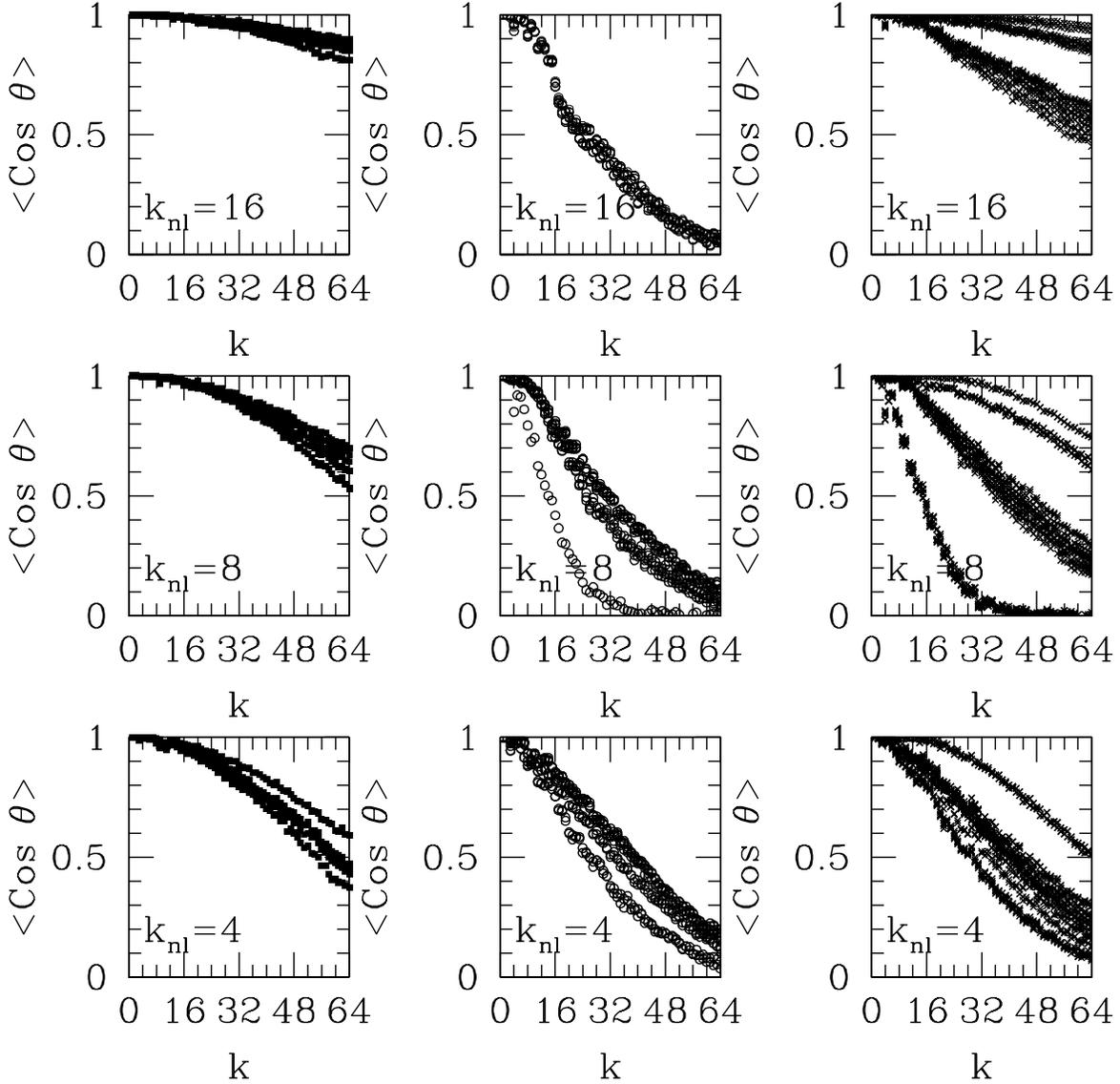,height=16cm}
\end{center}
\caption{
  The successive plots contain all the data for the averaged phase
  agreement between {\it all} of the simulations runs at the same
  stage; $< cos \theta > $ is defined in the text and is $1$ for
  agreement and $0$ for uncorrelated phases.  Individual functions are
  not distinguishable here, but it is clear that they fall into classes
  shown here.  The left column corresponds to all runs with good mass
  and force resolution and the same initial conditions.  The center
  column corresponds to anything cross-correlated with the run which
  continued the power-law perturbations to wave-numbers impossible for
  the HFLMR codes.  The right column contains everything else.  The
  high values found in the right column correspond to two HFLMR P$^3$M
  runs compared and two HFLMR Tree runs compared--which agree within
  the pair but not with anything else.
\label{fig:fig09}}
\end{figure}

Every correlation coefficient in Tables 2-4 now corresponds to a
function of $k$.  In view of the daunting task of trying to display
all these functions, we look instead for patterns.  Fortunately,
strong ones exist.  Figure~\ref{fig:fig09} includes all the data for
every run cross-correlated with every other run at the same stage.
All data points are plotted but not joined; many overlap.

They segregate in three groups (two of which begin to approach one
another at the last stage).  (1) The best group (solid squares,
left column)
corresponds to phase correlations among the group with PM, $k_{\rm c}
= 16k_{\rm f}$, the 128$^3$ and 64$^3$ \p3m runs and the 64$^3$ Tree
runs with the same initial conditions.  All runs which fit this
definition stay in this "high--agreement" group at all stages.  This
agreement deteriorates with time, albeit slowly.
Even at late times, $<\cos \theta>$ is close to 0.5 at the Nyquist
Frequency.  (2) The worst group (open
circles, middle column) corresponds to {\it anything} compared with the PM,
$k_{\rm c} = 64k_{\rm f}$ run.  Agreement with this worsens, then gets
better, as evolution proceeds.  (3) All other phase comparisons lie in
a third group (x's).  This includes comparisons between various HFLMR
runs, which worsen with time, and is shown in the right column.

This third group (x's) lies mostly between the other two, but some
comparisons are outliers, joining the top or bottom group.  This is
not consistent over evolution but a few regularities exist: The
outliers (x's) that join the high correlation group (squares) include
the two 32$^3$ \p3m runs compared and the two 32$^3$ Tree runs
compared.  Each of these pairs are strongly correlated but {\it do
not} correlate well with anything else, across code types or with
other runs with the same force resolution, but more particles, within
a code type.  The outlier x's that join the worst (circles) group
are, without exception, all comparisons involving one 32$^3$ particle
run and one with more particles within a given code, or two 32$^3$
runs from different code types.

These natural groupings correspond to those seen in the density
cross-correlations in Tables 2--4.  Given the close similarity of the
power spectrum in many pairs which cross-correlate badly, we conclude
that while differences in phases and amplitudes both exist, the phase
differences are primarily responsible for the correlation coefficient
differences.

Statistical measures that are sensitive to phase information will
clearly suffer more from discreteness that those that are not.  The
HFLMR strategy may produce a good approximation to the power spectrum
but not to anything that depends on phases unless the density field is
smoothed over the interparticle separation.

To summarize:  For given initial conditions, different codes agree
well only when the softening approaches the mean interparticle
separation. The absence of correct high frequency initial conditions
(center column) and discreteness errors (right column) prevent the
general agreement found in the left column.

\subsection{Displacement vs. Density Contrast}

For this series of tests we compare comoving displacement,
$|\vec{\Delta r}|$, between the PM particles and our standard series
of HFLMR runs as a function of the density contrast computed using the
PM particles.  We compute $|\vec{\Delta r}(\delta)|$ comparing to both
PM models ($k_{\rm c}=16k_{\rm f}$ and $64k_{\rm f}$) and at three
stages of evolution ($k_{\rm nl}=16k_{\rm f},$, $8k_{\rm f}$, and
$4k_{\rm f}$).

There are several trends of interest in these plots.  First, in
agreement with our results from the density cross-correlation tests
comparison with the PM $k_{\rm c}=16k_{\rm f}$ case appears to produce
better $|\vec{\Delta r}|$ than the PM $k_{\rm c}=64k_{\rm f}$ case.
Secondly, the trend of increasing agreement between the PM and HFLMR
codes as $\epsilon$ is increased also appears to hold here as well.
Thus, we see that increasing the number of particles present in the
HFLMR codes without changing the absolute force resolution causes
their time integration to converge more and more toward the PM code
result. In Figure 12, the best overall agreement with the PM
code is the P$^3$M code based on 128$^3$ particles.

\begin{figure}
\begin{center}
  \leavevmode\psfig{figure=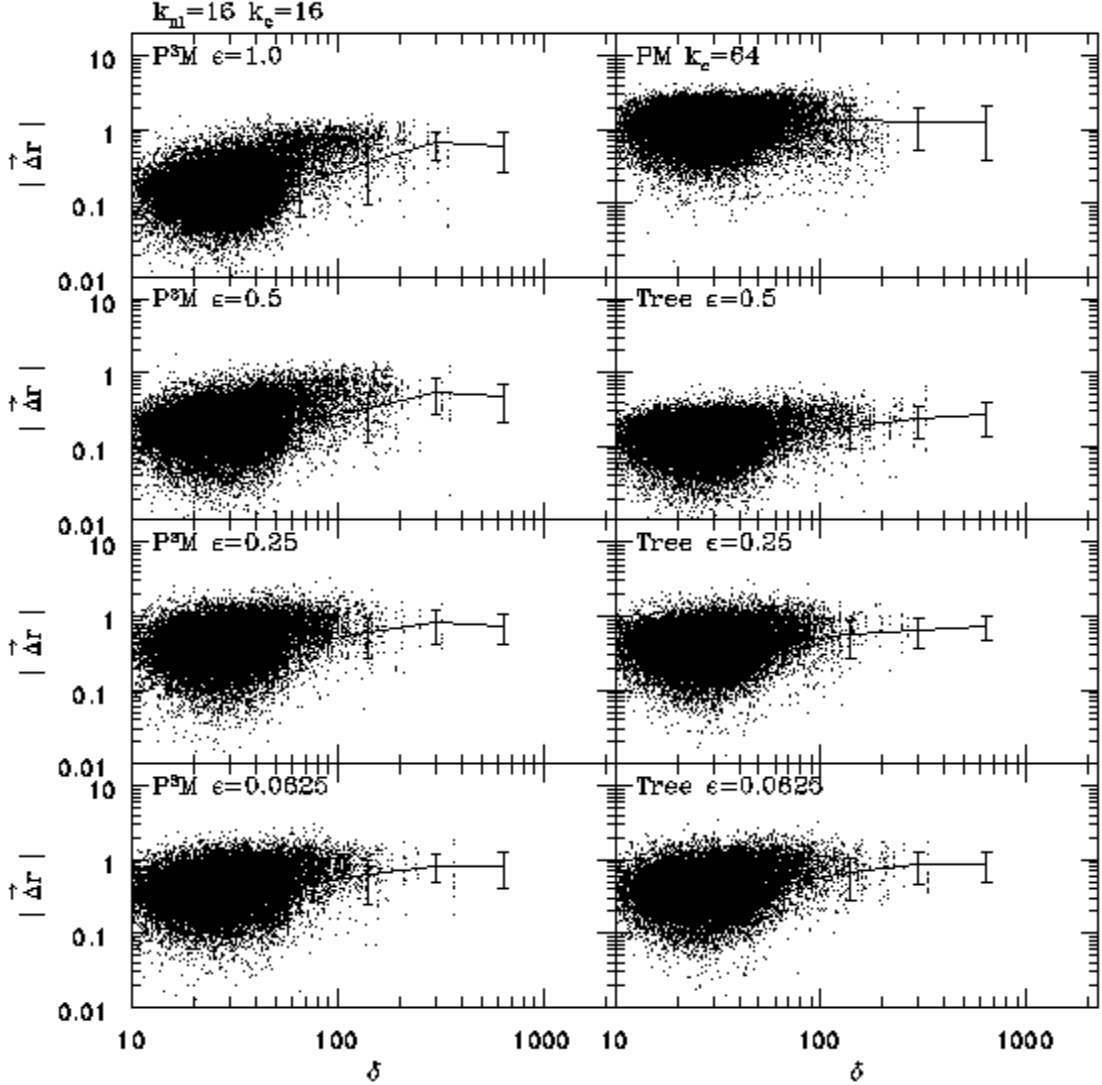,height=16cm}
\end{center}
\caption{
  The scatter plots contain the difference in positions between
  particles which had the same initial position and velocity in the PM
  run with $k_{\rm c} = 16k_{\rm f}$ and positions in other runs with
  identical initial conditions (and in the PM, $k_{\rm c} = 64k_{\rm
    f}$ run, upper right); The distance is plotted against the local
  density of the simulation in the fiducial PM run.  Shown here for
  the state $k_{\rm nl} = 16k_{\rm f}$.
  The line and error bars represnt the mean and standard deviation of
  the distance; the mean is sometimes not centered due to superposition
of dots.
\label{fig:rhofig10}}
\end{figure}

\begin{figure}
\begin{center}
  \leavevmode\psfig{figure=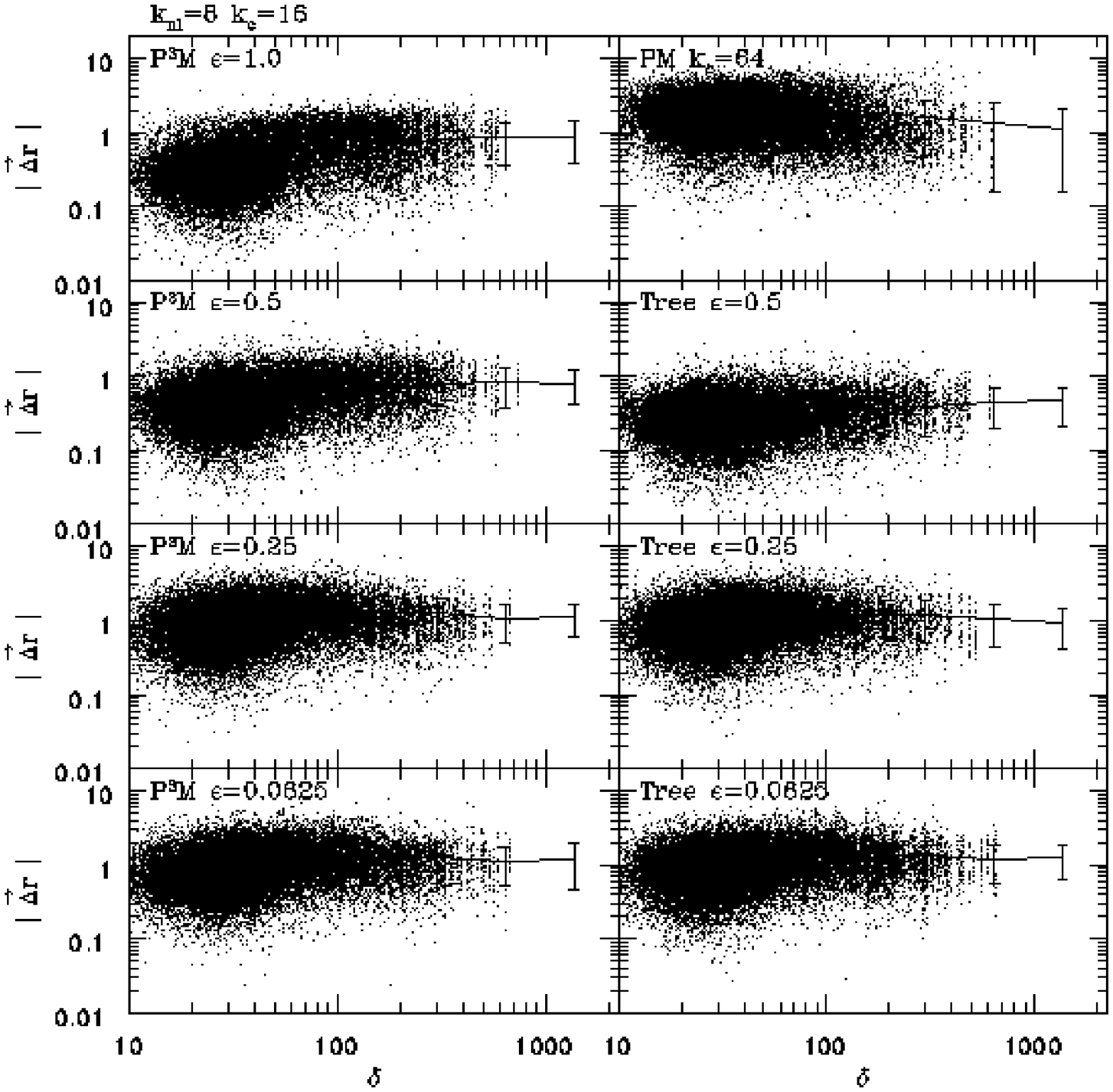,height=16cm}
\end{center}
\caption{
  The same as Fig. 10, except for the state $k_{\rm nl} = 8k_{\rm f}$.
\label{fig:rhofig11}}
\end{figure}

\begin{figure}
\begin{center}
  \leavevmode\psfig{figure=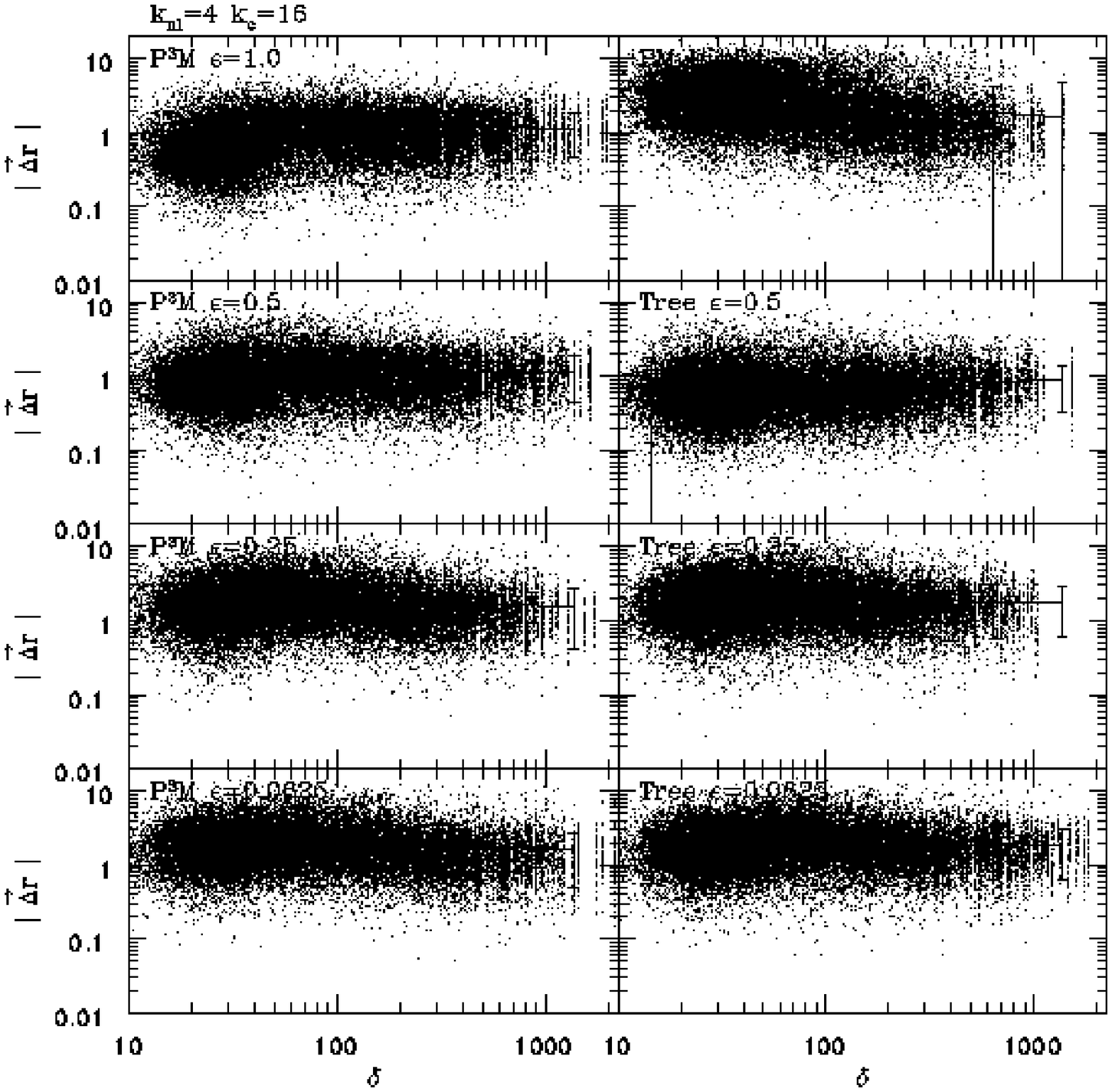,height=16cm}
\end{center}
\caption{
  The same as Fig. 10, except for the state $k_{\rm nl} = 4k_{\rm f}$.
\label{fig:rhofig12}}
\end{figure}

\begin{figure}
\begin{center}
  \leavevmode\psfig{figure=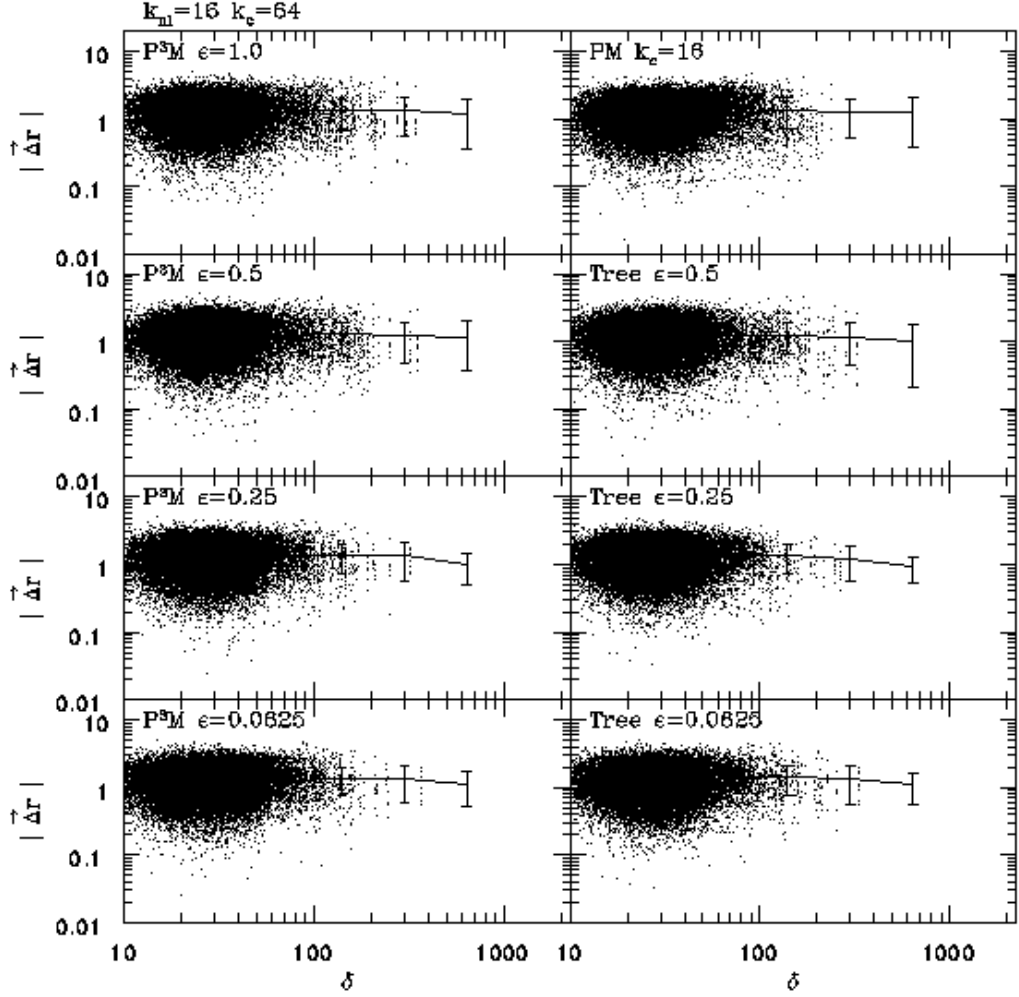,height=16cm}
\end{center}
\caption{
  The same as Fig. 10, except that the displacements are from the
  positions of the particles in the $k_{\rm c} = 64k_{\rm f}$ run, and
  the densities on the abscissa are drawn from that simulation.  The
  stage shown here is $k_{\rm nl} = 16k_{\rm f}$.
\label{fig:rhofig13}}
\end{figure}

\begin{figure}
\begin{center}
  \leavevmode\psfig{figure=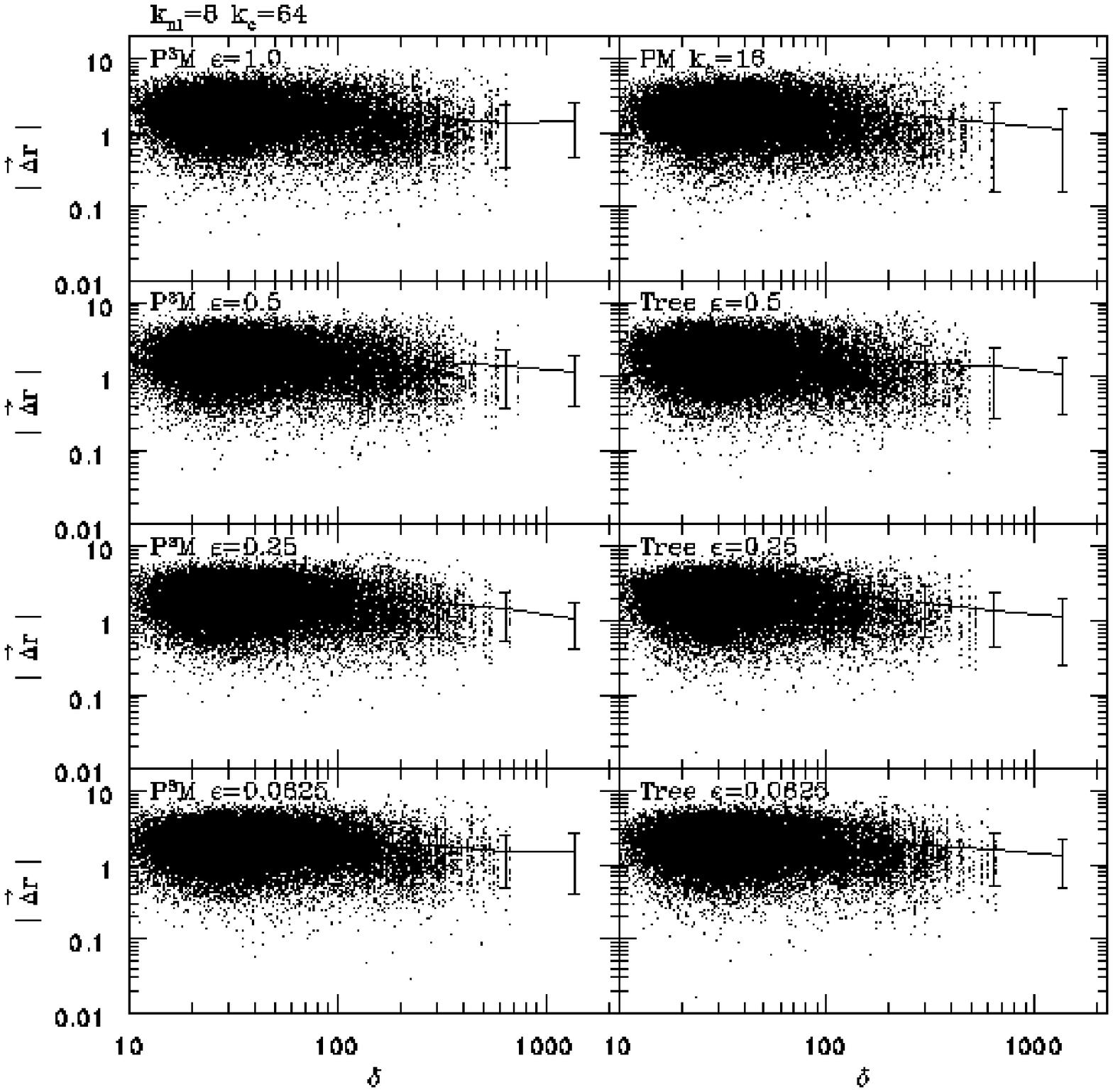,height=16cm}
\end{center}
\caption{
  The same as Fig. 13, except for the state $k_{\rm nl} = 8k_{\rm f}$.
\label{fig:rhofig14}}
\end{figure}

\begin{figure}
\begin{center}
  \leavevmode\psfig{figure=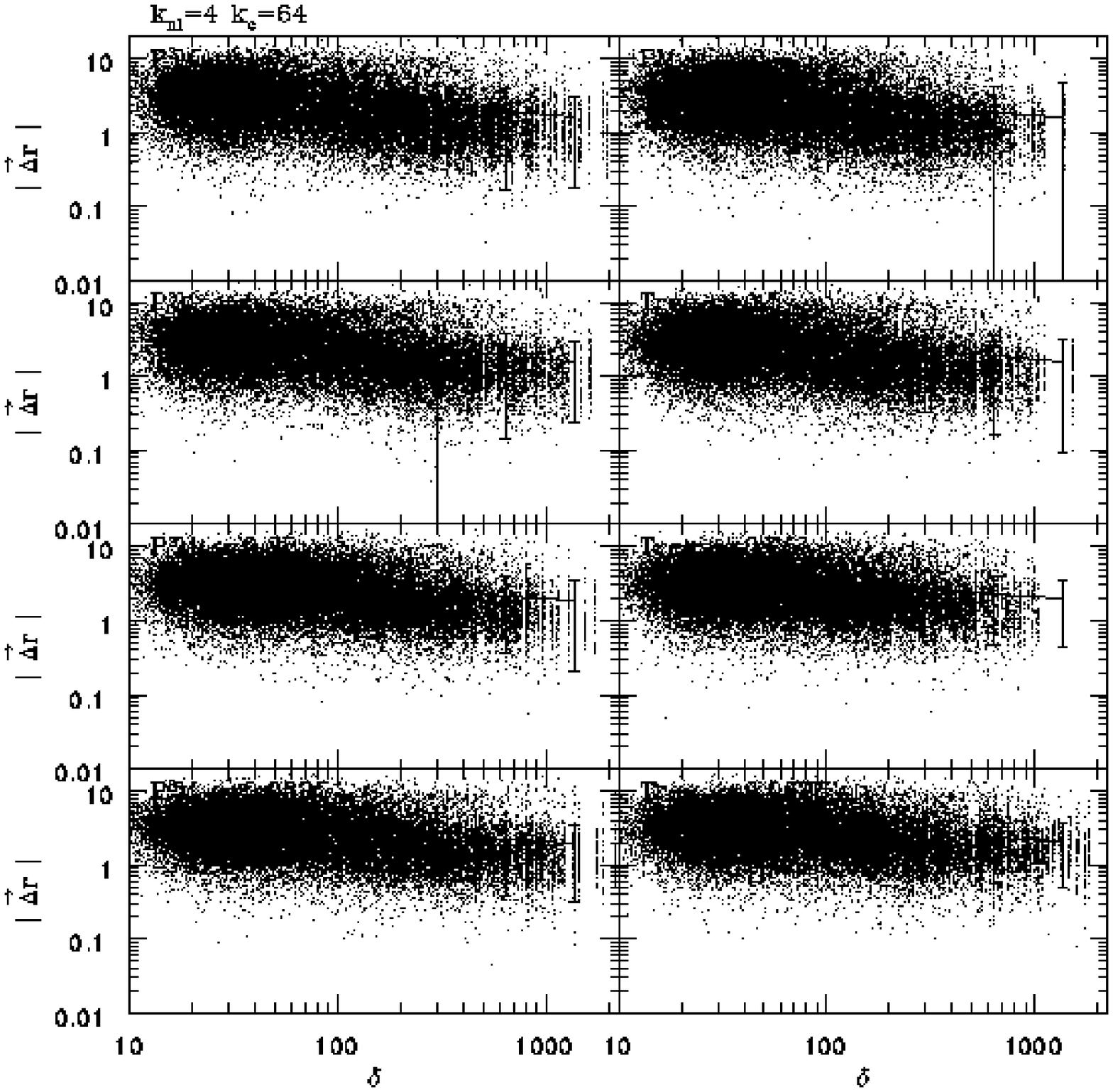,height=16cm}
\end{center}
\caption{
  The same as Fig. 13, except for the state $k_{\rm nl} = 4k_{\rm f}$.
\label{fig:rhofig15}}
\end{figure}

The differences that do exist between different codes with the same
initial conditions are amplified as the the particles continue to move
apart from one another.  Thus, comparing the top three panels of
Fig~\ref{fig:rhofig10}, Fig~\ref{fig:rhofig11}, and Fig~\ref{fig:rhofig12},
we see that amplification of errors is more rapid for smaller
$\epsilon$, even when the force resolution $a$ is held constant. In
Figures ~\ref{fig:rhofig13}, ~\ref{fig:rhofig14}, and ~\ref{fig:rhofig15}
the comparison is done against displacements of particles compared
with the PM $k_{\rm c}$ = 64$k_{\rm f}$ run.  In this case they are
uniformly bad.  Although the integration errors are greatly reduced by
having more particles, the absence of the extra power in the initial
conditions prevents convergence.  None of the runs, regardless of
its parameters, comes close.

\subsection{Cluster Analysis and Genus}

In cluster analysis, percolation properties are used to examine the
properties of connected regions.  Criteria of density thresholds or
the close approach of particles are used to decide which clumps are
connected.  This has been used to study the formation of structure in
simulations and in redshift surveys.  Percolation properties strongly
depend on sampling: the more particles in the model, the easier
percolation. Obviously percolation in the model corresponding to
Fig. 1b is easier than in Fig. 1a.

\begin{figure}
\begin{center}
  \leavevmode\psfig{figure=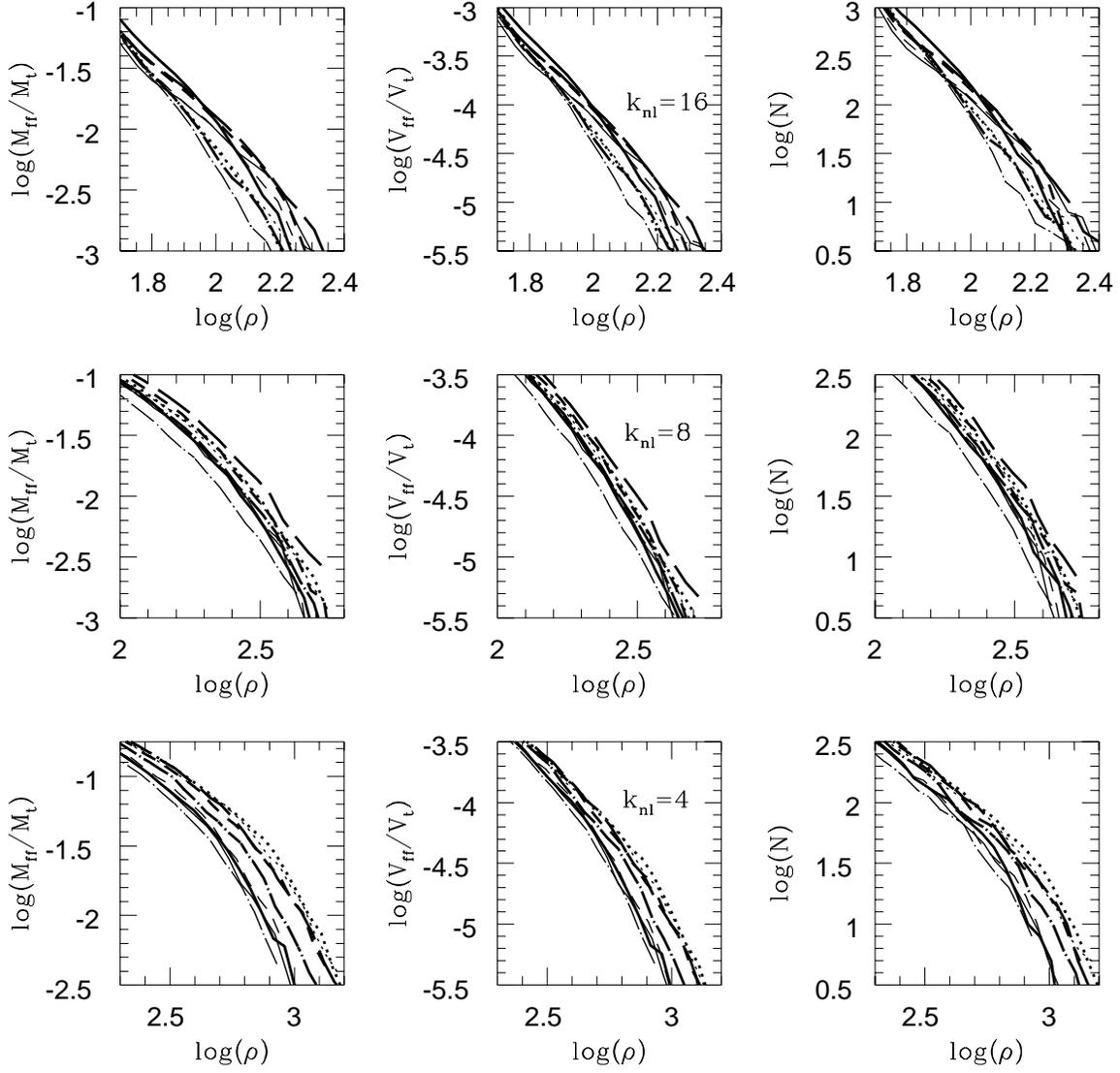,height=16cm}
\end{center}
\caption{
  Three rows show three stages in the evolution of the models. The
  left column shows the fraction of mass, the middle column - the
  fraction of volume, and the right column - the number of disjoint
  regions having densities greater than the density threshold shown on
  the horizontal. The densities were calculated on the $128^3$ mesh.
\label{fig:fig16}}
\end{figure}

\begin{figure}
\begin{center}
  \leavevmode\psfig{figure=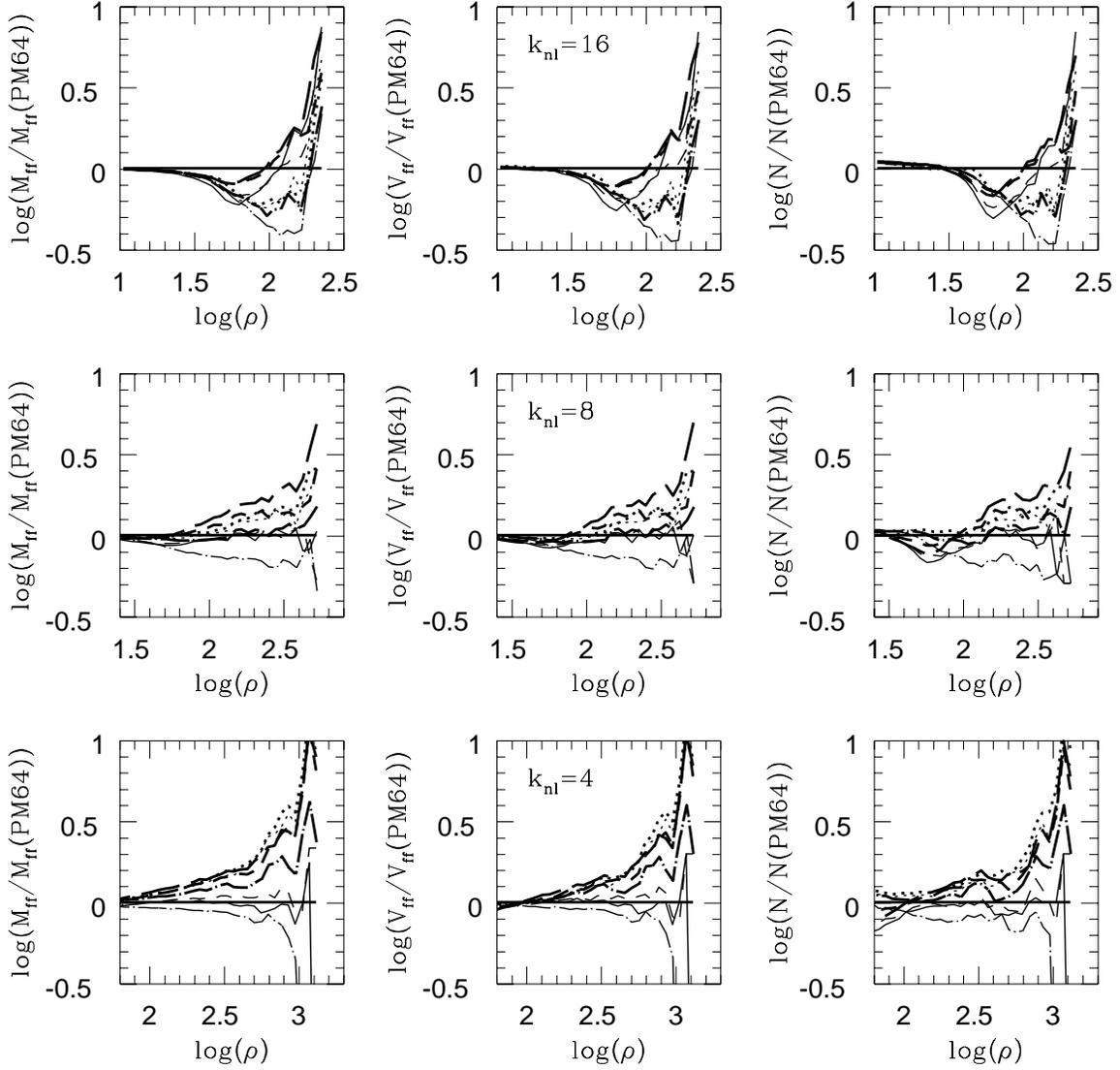,height=16cm}
\end{center}
\caption{
  The ratios of the same parameters plotted in Fig. 16 to the fiducial
  PM model with $N=128^3$, and $ k_c = 64$.
\label{fig:ratios}}
\end{figure}

There are less obvious differences in the number and other properties
of overdense regions. Fig.~\ref{fig:fig16} shows three statistics
calculated for the subsamples with equal number of particles ($N=
32^3$). The left panels show the fraction of mass in the regions
having densities above a given density threshold for all models at
three stages of evolution. The middle panels show the fraction of
volume occupied by these regions and the right panels show the total
number of overdense regions as functions of the density threshold. The
clumps were identified by using the percolation code described in Yess
\& Shandarin (1996). We show the high density part of these
distributions. The mass fraction runs from about 0.1\% to 10\% and the
number of clusters from 3 to 300 - 1000.  There are significant
differences between the models. Unfortunately, the patterns are
different at every stage therefore it is difficult to describe
them. The differences can be seen more clearly when the ratios are
plotted. To do so we choose PM($k_{\rm c} = 64k_{\rm f}$) as a
fiducial model and plot the logarithms of the ratios in
Fig.~\ref{fig:ratios}. We also extend the range of densities toward
lower densities in this figure so the convergence of the models can be
seen.

At the stage $k_{\rm nl} =16 k_{\rm f}$ all models converge at about
$\rho < 30$.  In the range $30 < \rho < 100 $ the fiducial PM64 model
has more clumps and more mass in clumps than any other model, however
at the very high densities $\rho > 100$ it has fewer clumps and less
mass in the clumps than any other model. The typical differences are 2
-3 times.  One could expect similar errors on small scales when
studying the early stages of clustering, such as Lyman-$\alpha$
absorbers or high-redshift galaxy formation, or the inner parts of dark matter
halos.

At the stage $k_{\rm nl} =8 k_{\rm f}$ the differences between models
are the least of all three stages.  All models except the Tree code
with $N=32^3, a=1, \epsilon=0.25$ have more mass in clumps; they are
also more numerous and occupy greater volume.  The \p3m model with
$N=128^3, a=1$ has the highest values of all statistics and the Tree
code with $N=32^3, a=1, \epsilon=0.25$ has the lowest. The \p3m and
Tree codes with $\epsilon=0.0625, a=0.25$ are significantly lower than
\p3m model with $N=128^3, a=1$.

\begin{figure}
\begin{center}
  \leavevmode\psfig{figure=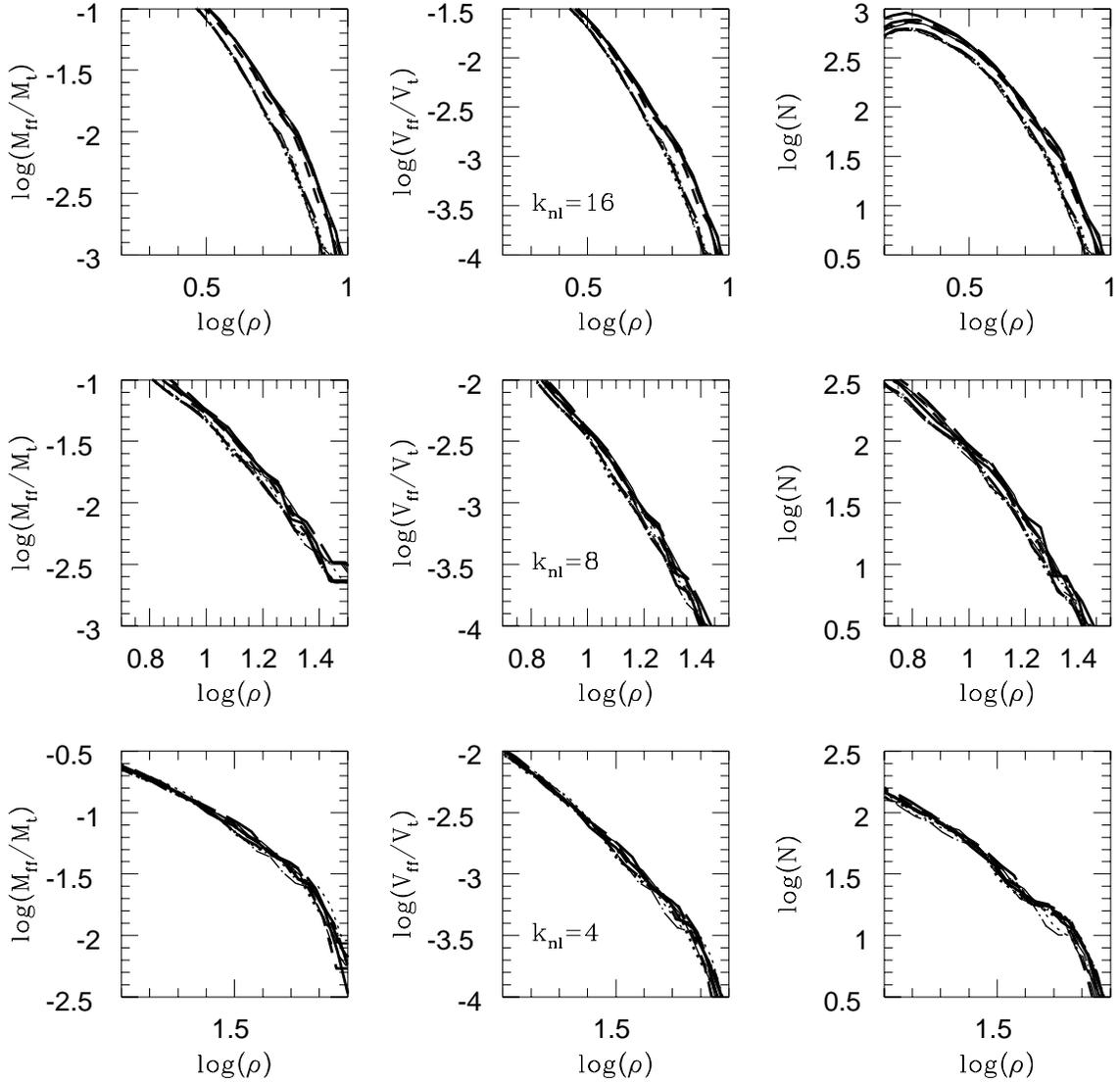,height=16cm}
\end{center}
\caption{
  Same as Fig. 16 except the densities were calculated on $32^3$ mesh.
\label{fig:fig18}}
\end{figure}

This pattern generally persists at the stage $k_{\rm nl} =4 k_{\rm
  f}$, however the amplitude of the differences is greater. The \p3m
model with $N=128^3, a=1$ is again among the highest but the \p3m and
Tree codes with best force resolution $\epsilon=0.0625, a=0.25$ are
now comparable to it. All the models converge at about $\rho < 100$.
When the densities are computed on the $32^3$ mesh the differences are
considerably smaller but still are noticeable at the $k_{\rm nl} =16
k_{\rm f}$ stage in Fig.~\ref{fig:fig18}. However, they almost vanish at
the final stage.  This variation of peak density with particle number
is completely consistent with results presented by Craig (1997) on the
central density of halos in N--body simulations. After this paper was
submitted, a preprint appeared (Moore et al. 1997) which shows the same effect
in CDM models at higher density levels.

\begin{figure}
\begin{center}
  \leavevmode\psfig{figure=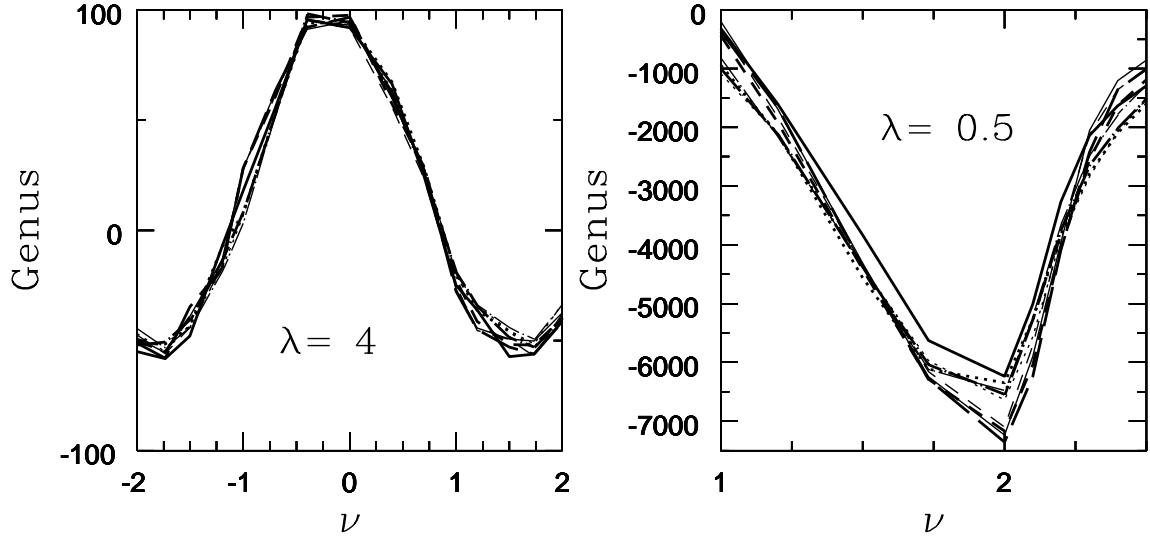,height=16cm}
\end{center}
\caption{
  The genus is plotted for the final stage $ k_{nl} = 4$: The left
  panel shows genus of the density fields smoothed on the scale of the
  mean particle separation; the right panel shows genus for the fields
  smoothed on half of the force resolution scale (only the part
  corresponding to high densities is shown).
\label{fig:fig19}}
\end{figure}

The genus is a measure of the connectivity of structures widely used
in cosmology.  It is specifically highly sensitive to phase
information.  All Gaussian distributions have the same distribution of
genus as a function of volume fraction, with the amplitude merely
being determined by an integral over the power spectrum.  We show the
genus here computed using the method of Weinberg et. al (1987). The genus
plotted for the last stage of the evolution in Fig.~\ref{fig:fig19}
is essentially in agreement with the other results: there are
differences up to 15\% between different codes on the scale of the
force resolution (right panel) and almost no differences on the scale
of the mean particle separation (left panel).  In fact with this
smoothing the N-body results are shown to reproduce well the
quasi-linear prediction for the genus curve. See Matsubara (1994) and
Matsubara \& Suto (1995) for details.

Cluster analysis has shown that if the clumps were selected using
isodensity surfaces calculated on the force resolution scale then
different codes would predict very different numbers of clumps, their
total volume and the total mass in the clumps. Differences as large as
factor of 2 or 3 are typical for the few hundred largest clusters. The
vertical scales in Fig.~\ref{fig:fig16} are about 2 -3 times greater
than the horizontal scales depending on the statistics and stage. This
means that if one reverses the question and asks what is the
difference in the density thresholds provided that same number of
clumps (or total mass) selected then the difference in densities would
be about 30 - 50\%. If the densities are computed on the scale of the
mean particle separation then all codes agree with each other at stage
$k_{\rm nl} =4 k_{\rm f}$ with about 20\% accuracy in terms of the
numbers, volumes and masses at given density thresholds or with
accuracy of 5\% in terms of density.  These differences are based on
clustering, not sampling, since we used the same number of particles
for each.  However, one can get an idea of the size of a possible
sampling effect by comparing the results shown here for PM with a
fully sampled PM density field using all 128$^3$ particles.  The
differences in these statistics between sparse and full PM samples are
much smaller than those we find between different codes.  Due to the
non-monotonic
nature of the changes we see with evolution, we cannot make any
predictions about how various codes and strategies would compare at
later stages of evolution.

\section{Discussion}

A number of the statistical results presented here show only moderate
differences between the various codes and our choices of $a$ and
$\bar{l}_{sep}$.  This is particularly true for the two--point
statistics, $P(k)$ and $\xi(r)$, which contain no phase
information. On the other hand, for $|<V_{12}(r)>|$ and
$\sigma_{V_{12}}(r)$ the dependence on $a$ and $\bar{l}_{sep}$ is
weak, while the dependence is strong for the phase correlations, the
mass density and number of clumps, as well as the phase correlations
and the density cross--correlations.  This implies that the
distribution of the phases may be the pivotal piece of information
that is needed to resolve this issue.  Unfortunately, no satisfactory
measure of the distribution of the phases exists.

We cannot easily compare our results with others, because no study
has included the variety of codes, mass resolution, and inclusion of
normally absent small--scale perturbations.  The most similar work is
the unpublished study coordinated by D.H. Weinberg.  An analysis
similar to our Figure 8 produced very similar results.  Efstathiou
et al. (1988) studied evolution of power--law initial conditions in
a P$^3$M code, using primarily low--order or averaged statistics.
They did find fluctuations of order 50\% in the rescaled multiplicity
function (their Figure 9); while not equivalent, this is roughly
compatible with our results in the central column of Figure 16.
It is interesting to keep in mind that in plasma physics (e.g.,
Hockney \& Eastwood 1988) and in beam dynamics (Habib 1997) it is
widely recognized that $\bar{l}_{sep} \ll a$ is a requirement to
accurately model a collisionless system.  This is in strong
contradiction to usual practice in cosmological simulations where
$\bar{l}_{sep} \gg a$. This is a survival in methodology from the
original cosmological N--body simulations which assumed to evolve the
distribution of galaxies and {\it not} the dark matter distribution
(in fact, they were generally called "galaxy clustering simulations"
at the time).  Note that we cannot appeal to which code produces the
best agreement with observations to settle our dilemma.  On small
scales, where we find significant differences, there is an unknown
correspondence between mass and light in the Universe.  This issue
must be settled by first principles, i.e., agreement on which
equations are to be solved and the most accurate methods to solve
them.

There is one crucial piece of evidence in our study.  No one would
dispute that the performance of any of our codes would become more
accurate as more particles are added, if the force resolution is held
constant. As is shown in many of our tests, this causes both Tree and
P$^3$M results (given the same initial conditions) to converge to our
PM run. For low-order statistics, such as the power spectrum and correlation
function, a small softening parameter appears to replicate the effect
of more particles (Although the results are not identical so it not
possible to determine how small the softening parameter can be made
without introducing artificial collisionality into the simulation).
But on the other hand, it is clear that decreasing the softening
parameter has profound effects on any measurement which depends
upon the phase information.

The sources of the observed errors are {\it not} random but systematic
in origin.  It is important to stress that there are two numerical
issues; (a) Two codes with identical initial conditions and similar
$a$ and $\bar{l}_{sep}$ should produce similar statistical results for
{\it all} statistical measures of the evolved particle distribution
and not just a subset.  Our results show a
convergence in the time integration between various codes as $a \sim
\bar{l}_{sep}$.  Unfortunately it is {\it not} standard in cosmology
(except in about half the PM applications) to use $a \sim
\bar{l}_{sep}$.  (b) The trouble is that the ``correct'' initial
conditions correspond to our case $k_{\rm c}=64k_{\rm f}$.  Thus, it
appears that the HFLMR codes are not able to correctly model the
$k_{\rm c}=64k_{\rm f}$ solution despite the past claims in the
literature and will produce solutions which only agree to
$\bar{l}_{sep}$ and not down to the force softening length $a$. This,
of course, creates a real dilemma when it comes to simulating truly
hierarchical structure. PM codes are fast, but the number of particles
are limited by disk storage requirements.  So--called adaptive mesh
refinement algorithms will suffer the same fate as the HFLMR methods
tested here since they are unable to input the true power spectrum
down to arbitrary scales initially.  This leaves us with nested--grid
PM methods (Villumsen 1988; Anninos, et al 1994; Splinter 1996) or related
strategies (Moore et al 1997) as a surviving
option.  These methods allow for both high force and
mass resolution simultaneously, thus guaranteeing that $a \sim
\bar{l}_{sep}$ is satisfied on the smallest scales and that the true
initial power spectrum is input {\it ab initio}. {\it Local} adaptive
mesh refinement schemes may offer some advantages if the refinement is
carefully designed and tested.  This may allow some reduction of the
integration error we have demonstrated, but cannot be a substitute for
inclusion of the proper initial conditions.

\bigskip
\bigskip

ALM and SFS wish to acknowledge the financial support of the
NSF--EPSCoR program, NASA grant N5-4039, and the National Center for
Supercomputing Applications. ALM also wishes to thank David Weinberg and P.J.E.
Peebles
for useful discussions as well as for an unpublished code comparison
project which proved useful.  RJS wishes to thank the Center for
Computational Sciences at the University of Kentucky for financial
support, and Salmon Habib for useful discussions.
 YS thanks RESCEU (Research Center for the Early Universe, University
of Tokyo), KEK (National Laboratory for High Energy Physics, Japan),
and the Astronomical Data Analysis Center of the National Astronomical
Observatory (where the Tree-code simulation were performed) for a
generous allocation of computer time. The research of YS was supported
by the Grants-in-Aid of the Ministry of Education, Science, Sports and
Culture of Japan No.07CE2002 to RESCEU, and by the Supercomputer
Project (No.97-22) of High Energy Accelerator Research Organization
(KEK).
All the authors thank an anonymous referee for comments which improved
the presentation.

%%%%%%%%%%
%Figure 1
%%%%%%%%%%

%%%%%%%%%%%%%%%%%%%%
%Figure 2
%%%%%%%%%%%%%%%%%%%%%
%%%%%%%%%%%%%%%%%%%%%%%%%%%%
%Figure 3
%%%%%%%%%%%%%%%%%%%%%%%%%%%%
%%%%%%%%%%%%%%%%%%%%%%%%%%%
%Figure 4
%%%%%%%%%%%%%%%%%%%%%%%%%%%%
%%%%%%%%%%%%%%%%%%%%%%%%%%%%%%%%%%
%Figure 5
%%%%%%%%%%%%%%%%%%%%%%%%%%%%%%%%%%
%%%%%%%%%%%%%%%%%%%%%%%%
%Figure 6
%%%%%%%%%%%%%%%%%%%%%%%%%%%%
%%%%%%%%%%%%%%%%%%%%%%%
%Figure 7
%%%%%%%%%%%%%%%%%%%%%%%%%%
%%%%%%%%%%%%%%%%%%%%%%%%%%%%%%%%
%Figure 8
%%%%%%%%%%%%%%%%%%%%%%%%%%%%%%%%%%
%%%%%%%%%%%%%%%%%%%%%%%%%%%%%%%%
%Figure 9
%%%%%%%%%%%%%%%%%%%%%%%%%%%%%%%%%

%%%%%%%%%%%%%%%%%%%%%%%%%%%%%%%%%%%%
%Figures 10 to 15
%%%%%%%%%%%%%%%%%%%%%%%%%%%%%%%%%%%
%%%%%
%%%%%%%%%%%
%%%%%%%%%%%%%%%
%%%%%%%%%%%%%%%%%%
%%%%%%%%%%%%%%%%%%%

%%%%%%%%%%%%%%%%%%%%%%%%%%%%%%%%

%%%%%%%%%%%%%%%%%%%%%%%%%%%%%%%%
%Figure 16
%%%%%%%%%%%%%%%%%%%%%%%%%%%%%%%%%

%%%%%%%%%%%%%%%%%%%%%%%%%%%%%%%%
%Figure 17
%%%%%%%%%%%%%%%%%%%%%%%%%%%%%%%%%

%%%%%%%%%%%%%%%%%%%%%%%%%%%%%%%%
%Figure 18
%%%%%%%%%%%%%%%%%%%%%%%%%%%%%%%%%

\clearpage

%%%%%%%%%%%%%%%%%%%%%%%%%%%%%%%%
%Figure 19
%%%%%%%%%%%%%%%%%%%%%%%%%%%%%%%%%

\clearpage

{\tiny
%%%%%%%%%%%%%%%%%%%%%%%%%%%%%%%%
%Table 1
%%%%%%%%%%%%%%%%%%%%%%%%%%%%%%%%
\begin{deluxetable}{ccccc}
\tablenum{1}
\tablecaption{Model Parameters for the Test Cases}
\tablehead{
\colhead{Code}&
\colhead{}&
\colhead{N}&
\colhead{$\bar{l}_{sep}$\tablenotemark{a}}&
\colhead{$\epsilon_{force}$\tablenotemark{b}}}
\startdata
PM      & $k_c=16$ & $128^3$ & 1.0 & 1.0    \nl
        & $k_c=64$ & $128^3$ & 1.0 & 1.0    \nl
P$^3$M  & $k_c=16$ & $32^3$  & 4.0 & 0.0625 \nl
        & $k_c=16$ & $32^3$  & 4.0 & 0.25   \nl
        & $k_c=16$ & $64^3$  & 2.0 & 0.5    \nl
        & $k_c=16$ & $128^3$ & 1.0 & 1.0    \nl
Tree    & $k_c=16$ & $32^3$  & 4.0 & 0.0625 \nl
        & $k_c=16$ & $32^3$  & 4.0 & 0.25   \nl
        & $k_c=16$ & $64^3$  & 2.0 & 0.5    \nl
\tablenotetext{a}{$\bar{l}_{sep}$ is the  mean particle separation  in
grid  cell  units.}
\tablenotetext{b}{$\epsilon_{force}$ is  in units of the mean particle
separation, $\bar{l}_{sep}$.   Note that  for  most of  the  runs $  a
\equiv  \epsilon_{force} \times \bar{l}_{sep} = 1$,  only for two does
$a=0.25$.}
\enddata
\end{deluxetable}

%%%%%%%%%%%%%%%%%%%%%%%%%%%%%%%%
%Table 2
%%%%%%%%%%%%%%%%%%%%%%%%%%%%%%%%
%\ptlandscape
\begin{deluxetable}{lccccccccccc}
\tablenum{2}
\tablecolumns{14}
\tablewidth{0pt}
\tablecaption{Cross--Correlations at $k_{nl}=16$}
\tablehead{
\colhead{}&
\multicolumn{2}{c}{PM}&
\colhead{} &
\multicolumn{4}{c}{P$^3$M}&
\colhead{} &
\multicolumn{3}{c}{Tree}\\
\cline{2-3} \cline{5-8} \cline{10-12}\\
\colhead{}&
\colhead{$k_c=16$} &
\colhead{$k_c=64$} &
\colhead{}&
\colhead{$128^3 \epsilon=1.0$}&
\colhead{$64^3  \epsilon=0.5$}&
\colhead{$\epsilon=0.25$} &
\colhead{$\epsilon=0.0625$} &
\colhead{}&
\colhead{$64^3 \epsilon=0.5$}&
\colhead{$\epsilon=0.25$} &
\colhead{$\epsilon=0.0625$} }
\startdata
%                             J16    J64      128p3m1 64p3m5 p3m25  p3m0625
%%64tree05 tree25 tree0625
PM ($k_c=16$)                 & ---  & 0.31 & & 0.94  & 0.93 & 0.77 & 0.76 & &
0.97 & 0.74 & 0.74 \nl
PM ($k_c=64$)                 & 0.92 & ---  & & 0.31  & 0.30 & 0.26 & 0.26 & &
0.31 & 0.26 & 0.26 \nl
P$^3$M $128^3 (\epsilon=1.0$) & 1.00 & 0.91 & & ---   & 0.95 & 0.72 & 0.72 & &
0.95 & 0.69 & 0.69 \nl
P$^3$M $64^3 (\epsilon=0.5$)  & 1.00 & 0.91 & & 1.00  & ---  & 0.81 & 0.81 & &
0.97 & 0.78 & 0.77 \nl
P$^3$M($\epsilon=0.25$)       & 0.99 & 0.91 & & 0.98  & 0.99 & ---  & 0.98 & &
0.81 & 0.96 & 0.96 \nl
P$^3$M($\epsilon=0.0625$)     & 0.98 & 0.91 & & 0.98  & 0.99 & 1.00 & ---  & &
0.81 & 0.95 & 0.95 \nl
Tree $64^3$ ($\epsilon=0.5$)  & 1.00 & 0.92 & & 1.00  & 1.00 & 0.99 & 0.99 & &
---  & 0.79 & 0.79 \nl
Tree($\epsilon=0.25$)         & 0.98 & 0.90 & & 0.98  & 0.98 & 1.00 & 1.00 & &
0.99 & ---  & 0.99 \nl
Tree($\epsilon=0.0625$)       & 0.98 & 0.90 & & 0.98  & 0.98 & 1.00 & 1.00 & &
0.99 & 1.00 & ---  \nl
\enddata
\end{deluxetable}

%%%%%%%%%%%%%%%%%%%%%%%%%%%%%%%%
%Table 3
%%%%%%%%%%%%%%%%%%%%%%%%%%%%%%%%
%\ptlandscape
\begin{deluxetable}{lccccccccccc}
\tablenum{3}
\tablecolumns{14}
\tablewidth{0pt}
\tablecaption{Cross--Correlations at $k_{nl}=8$}
\tablehead{
\colhead{}&
\multicolumn{2}{c}{PM}&
\colhead{} &
\multicolumn{4}{c}{P$^3$M}&
\colhead{} &
\multicolumn{3}{c}{Tree}\\
\cline{2-3} \cline{5-8} \cline{10-12}\\
\colhead{}&
\colhead{$k_c=16$} &
\colhead{$k_c=64$} &
\colhead{}&
\colhead{$128^3 \epsilon=1.0$}&
\colhead{$64^3  \epsilon=0.5$}&
\colhead{$\epsilon=0.25$} &
\colhead{$\epsilon=0.0625$} &
\colhead{}&
\colhead{$64^3 \epsilon=0.5$}&
\colhead{$\epsilon=0.25$} &
\colhead{$\epsilon=0.0625$} }
\startdata
%                             J16    J64      128p3m1 64p3m5 p3m25  p3m0625
%%64tree05 tree25 tree0625
PM ($k_c=16$)                 & ---  & 0.48 & & 0.87  & 0.84 & 0.60 & 0.58 & &
0.91 & 0.55 & 0.56 \nl
PM ($k_c=64$)                 & 0.94 & ---  & & 0.47  & 0.46 & 0.40 & 0.40 & &
0.46 & 0.37 & 0.38 \nl
P$^3$M $128^3 (\epsilon=1.0$) & 1.00 & 0.94 & & ---   & 0.88 & 0.59 & 0.57 & &
0.86 & 0.53 & 0.54 \nl
P$^3$M $64^3 (\epsilon=0.5$)  & 1.00 & 0.94 & & 1.00  & ---  & 0.65 & 0.63 & &
0.90 & 0.58 & 0.59 \nl
P$^3$M($\epsilon=0.25$)       & 0.97 & 0.92 & & 0.97  & 0.98 & ---  & 0.92 & &
0.64 & 0.86 & 0.87 \nl
P$^3$M($\epsilon=0.0625$)     & 0.96 & 0.92 & & 0.97  & 0.97 & 1.00 & ---  & &
0.62 & 0.83 & 0.84 \nl
Tree $64^3 (\epsilon=0.5$)    & 1.00 & 0.94 & & 0.99  & 1.00 & 0.98 & 0.98 & &
---  & 0.61 & 0.61 \nl
Tree($\epsilon=0.25$)         & 0.96 & 0.92 & & 0.96  & 0.96 & 0.99 & 0.99 & &
0.97 & ---  & 0.93 \nl
Tree($\epsilon=0.0625$)       & 0.96 & 0.92 & & 0.96  & 0.96 & 0.99 & 0.99 & &
0.97 & 1.00 & ---  \nl
\enddata
\end{deluxetable}

%%%%%%%%%%%%%%%%%%%%%%%%%%%%%%%%
%Table 4
%%%%%%%%%%%%%%%%%%%%%%%%%%%%%%%%
\begin{deluxetable}{lccccccccccc}
\tablenum{4}
\tablecolumns{14}
\tablewidth{0pt}
\tablecaption{Cross--Correlations at $k_{nl}=4$}
\tablehead{
\colhead{}&
\multicolumn{2}{c}{PM}&
\colhead{} &
\multicolumn{4}{c}{P$^3$M}&
\colhead{} &
\multicolumn{3}{c}{Tree}\\
\cline{2-3} \cline{5-8} \cline{10-12}\\
\colhead{}&
\colhead{$k_c=16$} &
\colhead{$k_c=64$} &
\colhead{}&
\colhead{$128^3 \epsilon=1.0$}&
\colhead{$64^3  \epsilon=0.5$}&
\colhead{$\epsilon=0.25$} &
\colhead{$\epsilon=0.0625$} &
\colhead{}&
\colhead{$64^3 \epsilon=0.5$}&
\colhead{$\epsilon=0.25$} &
\colhead{$\epsilon=0.0625$} }
\startdata
%                             J16    J64      128p3m1 64p3m5 p3m25  p3m0625
%%64tree05 tree25 tree0625
PM ($k_c=16$)                 & ---  & 0.65 & & 0.83  & 0.80 & 0.64 & 0.61 & &
0.84 & 0.52 & 0.49 \nl
PM ($k_c=64$)                 & 0.96 & ---  & & 0.63  & 0.63 & 0.57 & 0.55 & &
0.63 & 0.47 & 0.44 \nl
P$^3$M $128^3 (\epsilon=1.0$) & 0.99 & 0.95 & & ---   & 0.89 & 0.66 & 0.64 & &
0.84 & 0.50 & 0.48 \nl
P$^3$M $64^3 (\epsilon=0.5$)  & 0.99 & 0.96 & & 0.99  & ---  & 0.68 & 0.66 & &
0.84 & 0.49 & 0.47 \nl
P$^3$M($\epsilon=0.25$)       & 0.96 & 0.94 & & 0.97  & 0.97 & ---  & 0.86 & &
0.67 & 0.70 & 0.67 \nl
P$^3$M($\epsilon=0.0625$)     & 0.96 & 0.94 & & 1.00  & 1.00 & 1.00 & ---  & &
0.63 & 0.63 & 0.63 \nl
Tree $64^3 (\epsilon=0.5$)    & 1.00 & 0.96 & & 0.99  & 0.99 & 0.97 & 0.97 & &
---  & 0.60 & 0.56 \nl
Tree($\epsilon=0.25$)         & 0.92 & 0.90 & & 0.92  & 0.92 & 0.96 & 0.96 & &
0.94 & ---  & 0.87 \nl
Tree($\epsilon=0.0625$)       & 0.92 & 0.90 & & 0.91  & 0.91 & 0.96 & 0.96 & &
0.94 & 1.00 & ---  \nl
\enddata
\end{deluxetable}

}

\end{document}